\documentclass[lettersize,journal]{IEEEtran}
\usepackage{amsmath,amsfonts}
\newcommand{\attcolor}{black}
\usepackage[ruled,vlined,linesnumbered]{algorithm2e}
\usepackage{array}
\usepackage[caption=false,font=normalsize,labelfont=sf,textfont=sf]{subfig}
\usepackage{textcomp}
\usepackage{stfloats}
\usepackage{url}
\usepackage{verbatim}
\usepackage{graphicx}
\usepackage{cite}
\usepackage{pifont}
\usepackage{wrapfig}
\usepackage{amsmath}
\usepackage{amssymb}
\usepackage{mathtools}
\usepackage{amsthm}
\newtheorem{definition}{Definition}
\newtheorem{assumption}{Assumption}
\newtheorem{remark}{Remark}
\newtheorem{theorem}{Theorem}
\newtheorem{corollary}{Corollary}

\usepackage{multirow}
\usepackage{makecell}
\allowdisplaybreaks[4]
\usepackage{booktabs}
\usepackage{color}
\usepackage{hyperref}
\hypersetup{hypertex=true,
            linkcolor=blue,
            anchorcolor=blue,
            citecolor=blue}
\usepackage[table,xcdraw]{xcolor}

\begin{document}

\title{Unity is Power: Semi-Asynchronous Collaborative Training of Large-Scale Models with Structured Pruning in Resource-Limited Clients}

\author{Yan Li, Xiao Zhang, Mingyi Li, Guangwei Xu, Feng Chen, Yuan Yuan, Yifei Zou, Mengying Zhao, Jianbo Lu, Dongxiao Yu
\thanks{
\IEEEcompsocthanksitem This work was supported in part by  the National Natural Science Foundation of China under Grant 62202273 and Grant 62302247; 
in part by the Joint Key Funds of National Natural Science Foundation of China under Grant U24A20244; 
in part by Shandong Natural Science Foundation under Grant ZR2022QF140 and Grant ZR2025MS1105; 
in part by Key Research and Development Project of Shandong Province under Grant 2024CXPT052;
in part by Shandong Science Foundation for Excellent Young Scholars under Grant 2023HWYQ-007;
in part by the Fundamental Research Funds for the Central Universities.
 (\textit{Corresponding author: Xiao Zhang.})
\IEEEcompsocthanksitem Yan Li, Xiao Zhang, Mingyi Li, Guangwei Xu, Feng Chen, Yifei Zou, Mengying Zhao, Jianbo Lu, and Dongxiao Yu are with the School of Computer Science and Technology, Shandong University, Qingdao 266237, China. Email: yhyanli@mail.sdu.edu.cn, xiaozhang@sdu.edu.cn, limee@mail.sdu.edu.cn, gwxu@mail.sdu.edu.cn, fchen@mail.sdu.edu.cn, yfzou@sdu.edu.cn, zhaomengying@email.sdu.edu.cn, ziduke@163.com, dxyu@sdu.edu.cn.
\IEEEcompsocthanksitem Yuan Yuan is with the School of Software \& Joint SDU-NTU Centre for Artificial Intelligence Research (C-FAIR), Shandong University, Jinan 250100, China. Email: yyuan@sdu.edu.cn.}
}

\markboth{Journal of \LaTeX\ Class Files,~Vol.~14, No.~8, August~2021}%
{Shell \MakeLowercase{\textit{et al.}}: A Sample Article Using IEEEtran.cls for IEEE Journals}

\IEEEpubid{0000--0000/00\$00.00~\copyright~2021 IEEE}

\maketitle

\begin{abstract}
In this work, we study to release the potential of massive heterogeneous weak computing power to collaboratively train large-scale models  on dispersed datasets. 
  In order to improve both efficiency and accuracy in resource-adaptive collaborative learning, 
  we take the first step to consider the \textit{unstructured pruning}, \textit{varying submodel architectures}, \textit{knowledge loss}, and \textit{straggler} challenges simultaneously. 
We propose a novel semi-asynchronous collaborative training framework, namely ${Co\text{-}S}^2{P}$, with data distribution-aware structured pruning and cross-block knowledge transfer mechanism to address the above concerns. 
Furthermore, we provide theoretical proof that ${Co\text{-}S}^2{P}$ can achieve asymptotic optimal convergence rate of $O(1/\sqrt{N^*EQ})$. 
Finally, we conduct extensive experiments on two types of tasks with a real-world hardware testbed including diverse IoT devices.
The experimental results demonstrate that $Co\text{-}S^2P$ improves accuracy by up to 8.8\% and resource utilization by up to 1.2$\times$ compared to state-of-the-art methods, while reducing memory consumption by approximately 22\% and training time by about 24\% on all resource-limited devices.
\end{abstract}

\begin{IEEEkeywords}
 Collaborative Learning, Resource-Adaptive Learning, Structured Pruning.
\end{IEEEkeywords}

\section{Introduction}\label{sec:intro}
\IEEEPARstart{T}{he} pretrained foundation models (PFMs) with a huge amount of parameters, such as GPT-4~\cite{achiam2023gpt}, DeepSeek~\cite{liu2024deepseek} have witnessed great success in various fields, which demonstrates superior performance on kinds of downstream tasks,
such as CV~\cite{wu2023medical}, NLP~\cite{brown2020language}, robotics~\cite{cui2022can}, etc. 
However, training the large-scale pretrained models is becoming increasingly more expensive, usually requiring scaling to thousands of high-performance GPUs. 

In the real-world scenarios, heterogeneous weak computing power are commonly everywhere, such as mobile devices, equipped with limited computational and memory resources. 
Hence, it would be difficult and unaffordable for resource-constrained clients to train the full large-scale models. 
In addition, the resource-limited devices have heterogeneous data and bring 
"data silos" due to 
the promulgation of rigorous data regulations such as GDPR~\cite{hu2025federated,voigt2017eu,guo2024comprehensive,zhang2023federated}. 
Therefore, a practical problem arises: \textit{ 
How to release the potential of massive resource-limited devices to train large-scale models uniting the weak computing power and heterogeneous dispersed datasets collaboratively?
}


\IEEEpubidadjcol
\textcolor{\attcolor}{Traditionally, fruitful works have explored the lightweight collaborative training methods, such as quantization~\cite{ozkara2021quped, chen2024mixed, markov2023quantized}, compression~\cite{wu2024cg, jiang2022adaptive, niu2022federated}, distillation~\cite{gong2024federated, du2024decoupled, chen2023resource}, zeroth-order optimization~\cite{ling2024convergence,panchal2024thinking,qin2023federated} and pruning strategies~\cite{yu2021adaptive}.  
In addition, 
adaptively splitting the globally large models into submodels can facilitate the collaboration process in resource-limited federated learning (FL), such as RAM-Fed~\cite{wang2023theoretical}, PruneFL~\cite{jiang2022model}, OAP~\cite{zhou2022federated}, pFedGate~\cite{chen2023pFedGate}. }
However, when deploying the mentioned methods to prune the large-scale models 
\begin{figure}[t]
  \centering
  \includegraphics[width=0.5\textwidth]{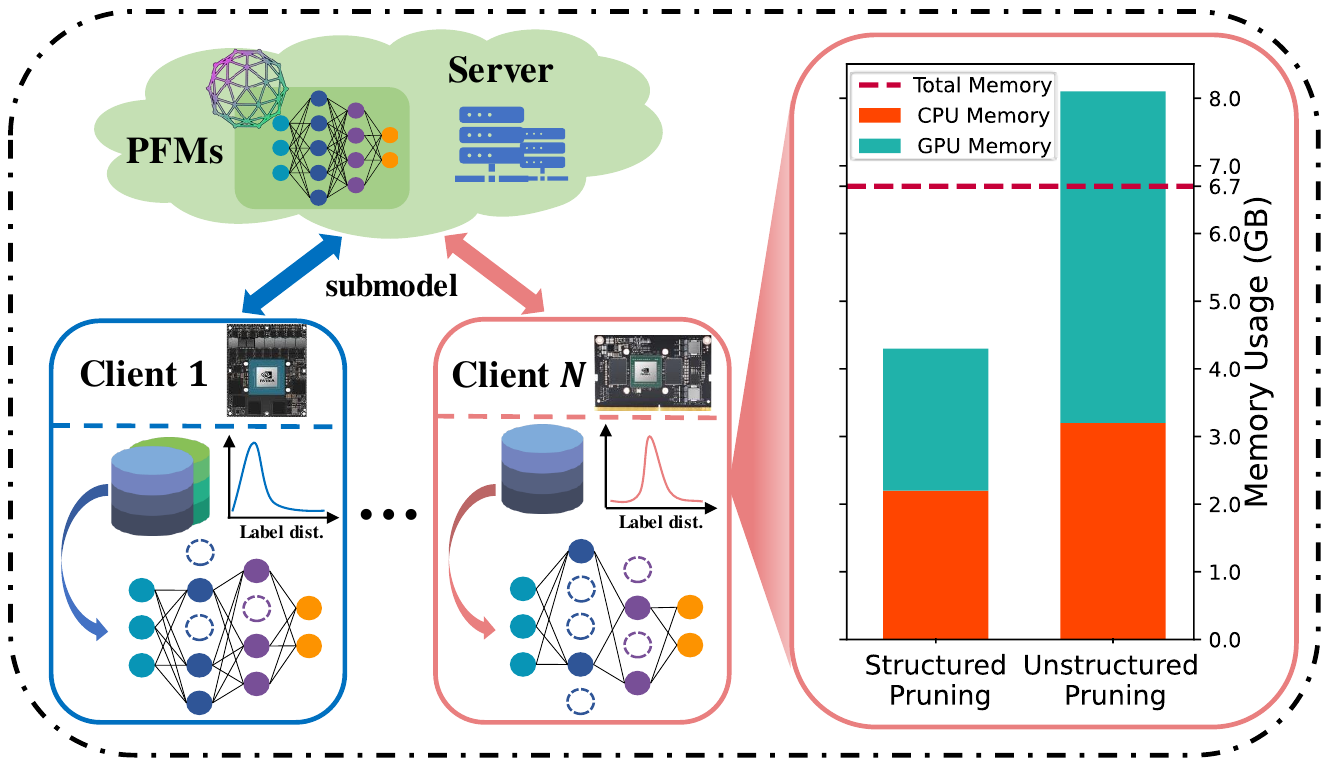}
  \caption{
  Unstructured pruning fails to train the splitted submodels due to memory constraints in the real-world resource-limited clients.}
  \label{motivation}
\end{figure}
in real-world resource-limited hardware, both \textit{efficiency} and \textit{accuracy} should be carefully considered,  
arising some tricky challenges. 
(1) \textit{Unstructured pruning.} 
The existing adaptive splitting methods adopt 
unstructured pruning~\cite{ilhan2023scalefl, isik2022sparse}, which only zeros out unimportant parameters without reducing the original parameter size.  
As shown in Fig.\ref{motivation}, unstructured pruning fails to train the splitted submodels due to memory constraints in real-world hardware. 
(2) \textit{Varying submodel architectures}. The randomized or weight-based submodels might not fit the heterogeneous data distributions in each client, which limits the learning potential of the generated submodels. 
(3) \textit{Knowledge loss}. 
The shallow submodels fail to learn high-level and complex knowledge due to the sparsity of the submodel architectures, 
causing knowledge loss in the resourced-limited clients. 
(4) \textit{Straggler}. 
The limited computation abilities of clients can cause enormous disparities in submodels training time, slowing down the large model's convergence and lowering down the utilization of dispersed computing power. 

Along this line, in this work, we propose a novel \underline{S}emi-asynchronous \underline{Co}llaborative training framework with \underline{S}tructured \underline{P}runing, named ${Co\text{-}S}^2{P}$, which can release the potential of massive heterogeneous resource-limited computing power to train a globally large model.  
In detail, ${Co\text{-}S}^2{P}$ designs a data distribution-aware structured pruning algorithm to ensure a balanced learning capability of submodels at both depth and width dimensions and accelerate the training process. 
The server first prunes the blocks of the large models in a rolling fashion  according to the available resources of clients. 
Then, the clients train the structured width-based segment-wise masks based on dispersed datasets to make the submodels fit the heterogeneous data distributions. 
Considering the knowledge loss of the pruned submodels 
in the resource-limited clients, we adopt 
self-distillation to implement cross-block knowledge transfer. 
To facilitate resource utilization and mitigate the straggler challenge, we design a semi-asynchronous aggregation strategy to further accelerate the large model convergence.  
Furthermore, we give a detailed convergence analysis of ${Co\text{-}S}^2{P}$, which can achieve an asymptotically optimal rate $O(1/\sqrt{N^*EQ})$.
Finally, we deploy ${Co\text{-}S}^2{P}$ on a real-world hardware testbed, in which 16 heterogeneous \textit{Jetson} devices can be united to train the vision transformer models with parameters up to 0.1 Billion,
demonstrating its superiority compared with the state-of-the-arts, 
while effectively reducing memory usage and training time.
Moreover, we validate the generality of ${Co\text{-}S}^2{P}$ across diverse tasks and heterogeneous IoT devices, demonstrating its broad applicability.
The main contributions are summarized as follows: 
\begin{itemize}
\setlength{\itemsep}{1pt}
\setlength{\parsep}{1pt}
\setlength{\parskip}{1pt}
    \item To release the potential of massive heterogeneous weak computing power to train large-scale models on dispersed datasets, we take the first step to improve both the efficiency and accuracy of the collaborative training by 
    considering the \textit{unstructured pruning}, \textit{varying submodel architectures}, \textit{knowledge loss}, and \textit{straggler} challenges into a united framework. 
    \item We propose a semi-asynchronous collaborative training framework ${Co\text{-}S}^2{P}$, in which the data distribution-aware 
    structured pruning is designed 
    at both depth and width dimensions while accelerating training. In addition, the structured pruned submodels are trained with self-distillation to implement cross-block knowledge transfer. 
    \item 
    We theoretically prove the strategy can converge with $O(1/\sqrt{N^*EQ})$, which shows that our semi-asynchronous aggregation strategy mitigates the straggler problem while balancing the convergence rate.
    \item 
    We conduct extensive experiments on a real-world hardware testbed, in which 16 Jetson devices can be united to train the large-scale models with parameters up to 0.1B.
    $Co\text{-}S^2P$ improves accuracy up to 8.8\% and resource utilization up to 1.2$\times$ compared with the SOTAs while reducing memory consumption and training time significantly. Furthermore, we demonstrate that ${Co\text{-}S}^2{P}$ can be scalable to different types of tasks.
\end{itemize}

\begin{figure*}[h]
    \center
    \includegraphics[width=0.9\linewidth]{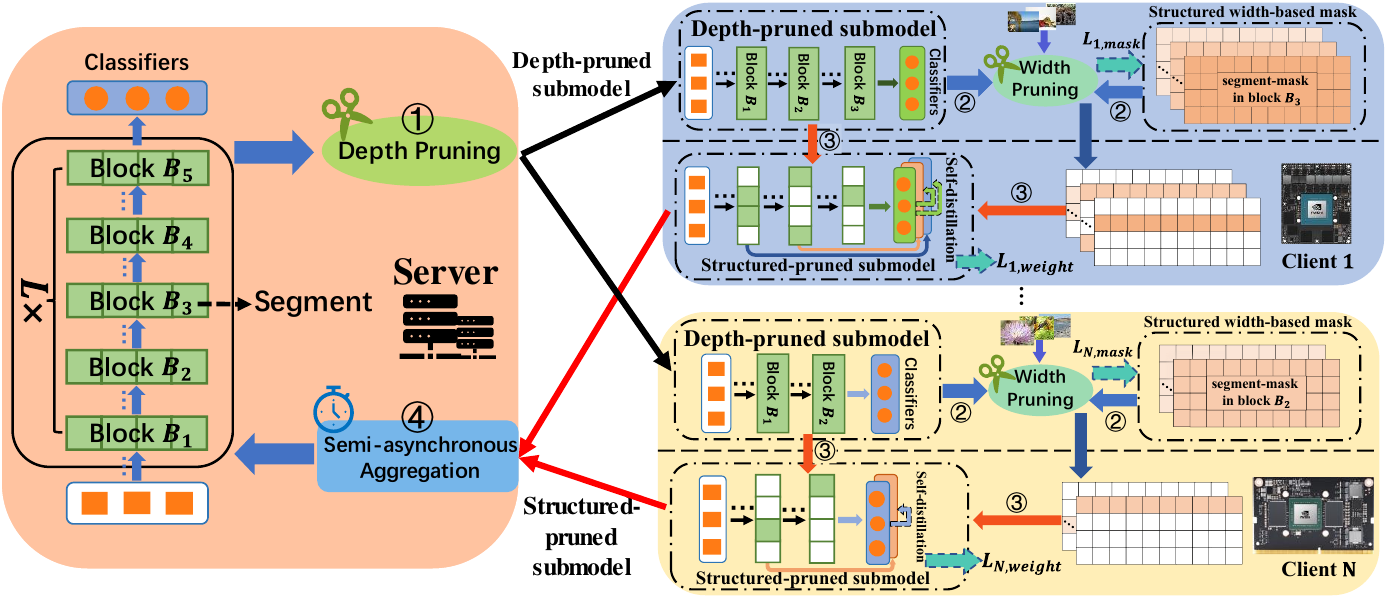}
     \caption{\textcolor{\attcolor}{Overview of $Co\text{-}S^2P$. 
     \textcircled{\raisebox{-1.0pt}{1}} The server prunes blocks at depth top-to-bottom and freezes the shallow blocks bottom-to-top according to the available resources of clients. 
     \textcircled{\raisebox{-1.0pt}{2}} The clients train structured width-based masks based on local datasets. 
     \textcircled{\raisebox{-1.0pt}{3}} The clients train strutrued-pruned submodels using self-distillation to implement cross-block knowledge transfer. 
     \textcircled{\raisebox{-1.0pt}{4}} We design a semi-asynchronous aggregation strategy to mitigate the problem of stragglers.}}
     \label{fig:overall_framework}
\end{figure*}

\section{Related Work}
\textbf{Lightweight Collaborative Training.} 
\textcolor{\attcolor}{
In recent years, fruitful works have been proposed for lightweight collaborative training such as quantization~\cite{ozkara2021quped, chen2024mixed, markov2023quantized}, compression~\cite{wu2024cg, jiang2022adaptive, niu2022federated},
distillation~\cite{gong2024federated, du2024decoupled, chen2023resource}
and zeroth-order optimization~\cite{ling2024convergence,panchal2024thinking,qin2023federated}. 
Specifically, FedMPQ~\cite{chen2024mixed} assigns different quantization bit-widths to local models via quantization regularization. 
For compression, PriSM~\cite{niu2022federated} decomposes all layers into kernels, and selects a subset to construct submodels. 
However, such methods are ineffective for large-scale Transformer-based models~\cite{vaswani2017attention}. 
FedIOD~\cite{gong2024federated} proposes a data-free FL framework based on local-to-central collaborative distillation but imposes prohibitive teacher‑model overhead. 
FedMeZO~\cite{ling2024convergence} and FedKSeed~\cite{qin2023federated} estimate gradients by forward propagation with zeroth-order optimization in federated learning scenarios.}

\textcolor{\attcolor}{
In addition, several works introduce  pruning~\cite{ilhan2023scalefl,jiang2022model,yu2021adaptive,wang2024save,bai2024fedspallm} and masking strategies~\cite{li2021fedmask,isik2022sparse,chen2023pFedGate,he2023gluefl} to construct resource-adaptive submodels for resource-limited clients. 
For instance, PruneFL~\cite{jiang2022model}, GLUEFL~\cite{he2023gluefl} and FedSpaLLM~\cite{bai2024fedspallm} unstructuredly prune by evaluating parameter importance, overlooking the effects of heterogeneous data. 
FedCyBGD~\cite{wang2024save} utilizes cycle block gradient descent and designs a hybrid granularity pruning scheme to periodically update the model.
RAM-Fed~\cite{wang2023theoretical} randomly generates unstructured masks and constructs submodels for clients, which limits convergence performance. 
Furthermore, these unstructured pruning and masking strategies cannot reduce the memory overhead and accelerate training, as shown in Fig.\ref{motivation}. 
To enhance both efficiency and accuracy, we propose the data distribution-aware structured pruning algorithm and self-distillation mechanism, fitting limited resources and improve the efficiency of knowledge transfer and convergence.}

\textbf{Asynchronous Collaborative Training.} 
\textcolor{\attcolor}{In the heterogeneous collaborative training scenarios, 
it is common to have slower nodes (stragglers) that are the bottleneck of each training round, leading to significant idle computing time.
Fully asynchronous methods~\cite{li2025accelerating,xie2019asynchronous, zhu2023robust} address this by enabling the server to immediately update the global model whenever it receives a local model.
FedASMU~\cite{liu2024fedasmu} proposes an asynchronous FL system with the dynamical model aggregation and adaptive local model adjustment method.
However, the fully asynchronous methods cause decrease in convergence speed and performance due to excessive update staleness.}

\textcolor{\attcolor}{Additionally, the semi-asynchronous collaborative training works use buffering~\cite{nguyen2022federated,zhang2023timelyfl} or clustering~\cite{chai2021fedat,miao2023robust} to mitigate the straggler problem while increasing convergence speed.
FedBuff~\cite{nguyen2022federated} sets an update buffer on the server side and perform aggregation only after a predefined number of client updates have been buffered. 
FedAT~\cite{chai2021fedat} synergistically integrates synchronous intra-tier training with asynchronous cross-tier training.
In our work, we design a novel semi-asynchronous aggregation strategy to accelerate collaborative training while mitigating the impact of stale clients.} 


\begin{table}[t]
 \caption{\textcolor{\attcolor}{Frequently used notations and corresponding descriptions}}
 \label{table:notation}
 \centering
	\begin{tabular}[t]{p{1.3cm} p{6.2cm}}
		\hline
		\textcolor{\attcolor}{Notations} & \textcolor{\attcolor}{Descriptions} \\ 
		\hline
            \textcolor{\attcolor}{$w$} & \textcolor{\attcolor}{the parameters of global model}\\
            \textcolor{\attcolor}{$w_n$} & \textcolor{\attcolor}{the parameters of submodel for client $n$}\\
            \textcolor{\attcolor}{$R_n$} & \textcolor{\attcolor}{the pruning rate for client $n$}\\
            \textcolor{\attcolor}{$w_n^{l_i}$} & \textcolor{\attcolor}{the parameters of $l_i$-th layer of submodel for client $n$}\\
            \textcolor{\attcolor}{$M_n^{l_i}$} & \textcolor{\attcolor}{the binary mask for structured pruning} \\
            \textcolor{\attcolor}{$f_n(w;x,y)$} & \textcolor{\attcolor}{the loss function for client $n$} \\
            \textcolor{\attcolor}{$N_{q}^i$} & \textcolor{\attcolor}{the set of clients training segment $i$ in round $q$}\\
            \textcolor{\attcolor}{$N^*$} & \textcolor{\attcolor}{minimum covering number: $N^*=\min\limits_{q,i}|N_{q}^i|$}\\
            \textcolor{\attcolor}{$\Delta_{q}^i$} & \textcolor{\attcolor}{the accumulated updates for parameter $i$ of global model in round $q$}\\
            \textcolor{\attcolor}{$\Delta_{q,n}$} & \textcolor{\attcolor}{the accumulated local updates from client $n$ on itself submodel in round $q$}\\
            \textcolor{\attcolor}{$\hat{\eta}$} & \textcolor{\attcolor}{the learning rate for training mask}\\
            \textcolor{\attcolor}{$\eta$} & \textcolor{\attcolor}{the learning rate for training submodel}\\
            \textcolor{\attcolor}{$\lambda_1$} & \textcolor{\attcolor}{the extent to constraint submodels}\\
            \textcolor{\attcolor}{$\lambda_2$} & \textcolor{\attcolor}{the balance of cross entropy loss and self-distillation}\\
            \textcolor{\attcolor}{$\tau_{q,n}$} & \textcolor{\attcolor}{the delay rounds that client $n$ has not taken part in global aggregation} \\
            \textcolor{\attcolor}{$\mu$} & \textcolor{\attcolor}{the aggregation minimal ratio}\\
            \textcolor{\attcolor}{$T_{clk}$} & \textcolor{\attcolor}{the waiting interval}\\
		\hline
	\end{tabular}
\end{table}

\vspace{-1em}
\section{Problem Formulation}
In this paper, we aim to train globally large models collaboratively 
while optimizing the utilization of diverse resource-limited computing power.  
Specifically, 
we assume there exist $N$ clients with different resources and locally heterogeneous datasets  $\mathcal{D}_n=\left\{\left(\mathbf{X}_i, y_i\right)\right\}_{i=1,\dots,|\mathcal{D}_n|}$, where $\left(\mathbf{X}_i, y_i\right)$ denotes the $i$-th training data sample and its ground truth label, and $n \in$ $\{1, \ldots, N\}$. 
In the resource-limited collaborative learning scenario, 
the optimization function is as follows:
\begin{equation}\label{optim}
\min _{w \in \mathbb{R}^d} F(w) := \sum_{n=1}^N \frac{1}{N} \cdot \mathbb{E}_{(x, y) \sim \mathcal{D}_n}\left[f_n\left(w; x, y\right)\right],
\end{equation}
where $w$ and $w_n$ are the parameters of global model and pruned submodels respectively. $f_n\left(w; x, y\right) \triangleq \ell\left(w_n(x), y\right)$, where $\ell(\cdot)$ is the loss function. \textcolor{\attcolor}{The frequently used notations is listed in Tab.\ref{table:notation}.}

\section{\textit{$Co\text{-}S^2P$} Framework Design}\label{sec:framewrok_design}
\textcolor{\attcolor}{We propose a semi-asynchronous collaborative training framework ${Co\text{-}S}^2{P}$ to \textit{release the potential of massive resource-limited devices}, as shown in Fig.\ref{fig:overall_framework}.
In ${Co\text{-}S}^2{P}$, we design a data distribution-aware structured pruning algorithm to ensure a balanced learning capability of submodel at both the depth and width dimensions while accelerating training (Sec.\ref{DDALA}). 
The server first prunes blocks in a rolling fashion according to the available resources of clients. 
Then, in order to make submodels fit the heterogeneous data distributions, the clients train structured width-based segment-wise masks based on local datasets.
Considering the pruning-induced knowledge loss in resource-limited clients, we train submodels with self-distillation to implement cross-block knowledge transfer within the same submodel (Sec.\ref{CBKT}). 
Finally, to mitigate the straggler problem, we design a pruning-based semi-asynchronous aggregation strategy (Sec.\ref{SAS}). }

\begin{figure}[t]
\centering
  \includegraphics[width=0.47\textwidth]{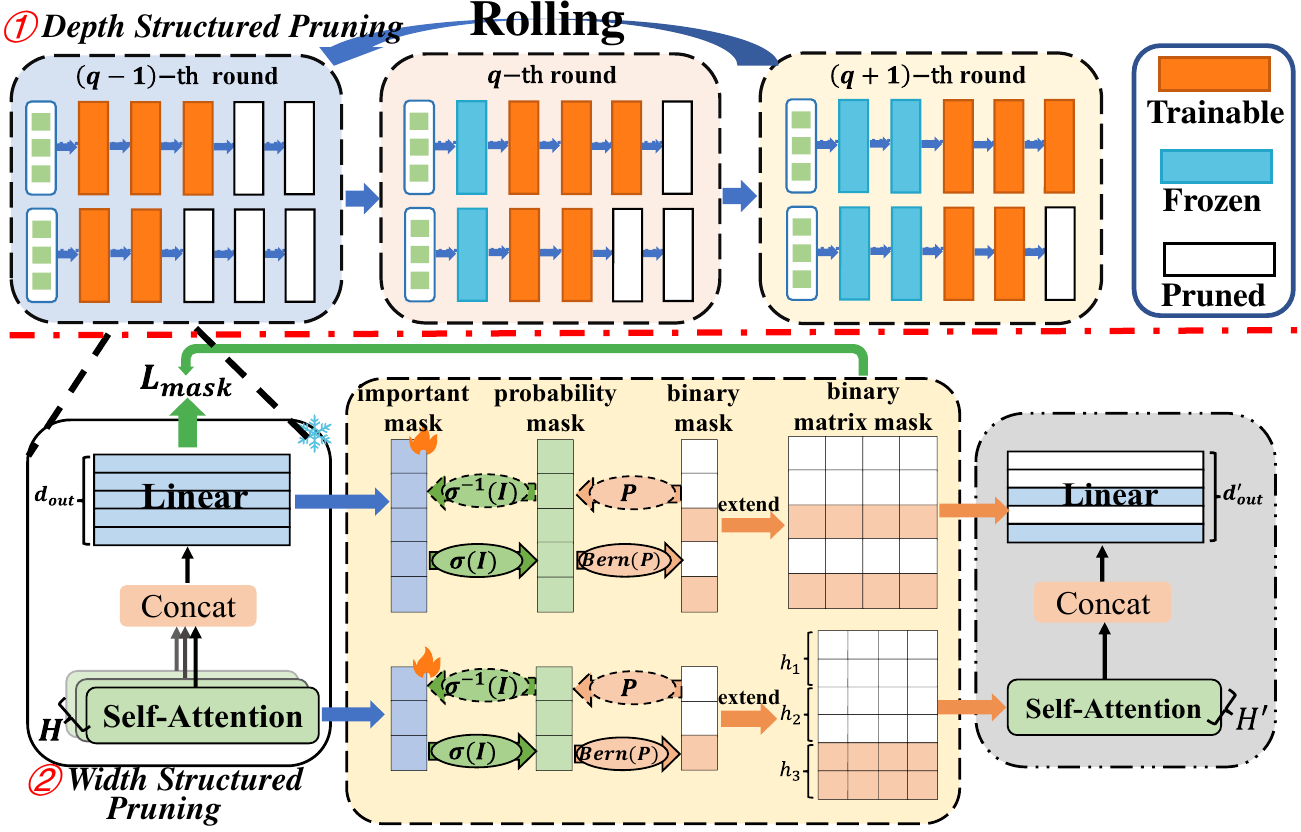}
  \caption{\textcolor{\attcolor}{Details of data distribution-aware balanced structured pruning in both depth and width dimensions. In the depth dimension, the server prune the Transformer blocks top-to-bottom and freeze the blocks bottom-to-top. Subsequently, the heterogeneous clients prune the depth-pruned submodel by training the width-based structured masks based on the local data distribution.} }
  \label{fig:mask_strategy}
\end{figure}

\subsection{Data Distribution-Aware Structured Pruning}\label{DDALA}
In this section, we focus on the balanced structured pruning of the globally large models at both depth and width dimensions, as shown in Fig.\ref{fig:mask_strategy}. 
For the heterogeneous clients with diverse limited resources and dispersed datasets, we measure the training time for a round using different submodels with different capacities and obtain a proper pruned rate $R_n$ for each client according to the available resources, which consists of the width pruned rate $R_n^{width}$ and the depth pruned rate $R_n^{depth}$.
The pruned rate is defined as the ratio of the structured-pruned model size of the full model size. By constraining these two values, the submodels can obtain balanced learning capacities.

However, the traditional depth pruning method has a strong assumption that some of the available clients can afford to train the full model~\cite{ilhan2023scalefl}, i.e., there exists at least one client $n^*$ with $R_{n^*}^{depth}=1$. 
We design a novel deep pruning mechanism to break the constraints and ensure that the whole blocks can be trained as evenly as possible by all the clients. For client $n$, the server first generates the depth-pruned submodel with $\lfloor L\times R_n^{depth} \rfloor$ consecutive trainable blocks by freezing the shallow blocks bottom-to-top and pruning the deep blocks top-to-bottom in a rolling fashion.

In order to make the submodels fit the heterogeneous data distributions, the clients train the width-based structured masks based on the local data distributions.
A Transformer-like model mainly consists of linear layers and Multi-head Self-Attention (MSA) layers. Consequently, we design trainable segment-wise masks for two types of layers, where the segment represents part parameters of linear layer or head in MSA layer. 
We denote the weight matrix and its binary mask by $w_n^{l_i} \in \mathbb{R}^{d_{out}\times d_{in}}$ and $M_n^{l_i} \in \{0,1\}^{d}$ respectively, where $l_i$ represents the $i$-th layer. 
For the linear layer, each value of $M_n^{l_i} \in \{0,1\}^{d_{out}}$ indicates whether the part of the layer is pruned or not. 
For the MSA layer, each value of $M_n^{l_i} \in \{0,1\}^{h}$ indicates whether the head is pruned or not and $h$ represents the number of heads.

However, considering that the segment-wise mask $M_n^{l_i}$ is a binary tensor, directly applying the existing back-propagation methods to model training would lead to gradients backward errors. 
Inspired by~\cite{zhou2019deconstructing, isik2022sparse}, we introduce the corresponding segment-wise importance mask $I_n^{l_i} \in \mathbb{R}^{d}$ and segment-wise probability mask $P_n^{l_i} \in [0,1]^{d}$. 
To enable back-propagation, we apply the sigmoid function scaling to the importance mask, and further use the Bernoulli distribution~\cite{coolidge1925introduction} to generate the binary mask $M_n^{l_i} \sim  Bern(\frac{1}{1+e^{-I_n^{l_i}}})$.
Then we use the inverse of the sigmoid function and record $P_n^{l_i}$ with the straight-through estimator~\cite{bengio2013estimating} for back-propagation.
\textcolor{\attcolor}{In the algorithm, the segment-wise importance mask $I^l_n$ and its corresponding probability mask $P_n^l$ that are learned jointly with parameter weights. 
Each entry of $I_n^l$ is updated by back-propagation so its magnitude continuously reflects how critical the corresponding coupling structures (heads in MSA or segments in linear layers) is for local data distribution.}

\textcolor{\attcolor}{Furthermore, the binary matrix mask $\hat{M}_n^{l_i}  \in \mathbb{R}^{d_{out}\times d_{in}}$ is introduced as an extension of $M_n^{l_i}$ to produce Hadamard product with the weight matrix of the model. 
For training the personalized mask $M_n^{l_i}$ efficiently, we freeze the corresponding weights of the depth-pruned submodels.  
Considering the resource-limited problem, we use the submodel size to reflect computation and communication costs. The loss function is designed to balance two important goals: submodel performance and training efficiency under resource constraints. as follows:}
\begin{equation}
\begin{aligned}
\label{eq:mask_loss}
\mathcal L_{n, mask} = \mathcal L_{CE}(\hat{y}, y) + \lambda_1|\frac{\sum_{l_i \in S}sum(\hat{M}_n^{l_i})}{\sum_{l_i \in S}size(\hat{M}_n^{l_i})}-R_n^{width}|,
\end{aligned}
\end{equation}
where $\mathcal L_{CE}(\hat{y}, y)$ represents the cross entropy loss for target class $y$, $S$ represents the set of the linear layers and the MSA layers. $\lambda_1$ is the hyperparameter that indicates the extent to which the submodel adapts to the limited resource. For each client, $\frac{\sum_{l_i \in S}sum(\hat{M}_n^{l_i})}{\sum_{l_i \in S}size(\hat{M}_n^{l_i})}$ represents the proportion of non-zero values in $M_n^{l i}$, that is, the actual width-pruning rate.
\textcolor{\attcolor}{In summary, the loss is designed to achieve a balance between the model performance and the efficiency of pruning. This dual-objective approach enables resource-efficient model training without sacrificing performance in resource-limited scenarios.}

Finally, we perform Hadamard product for the depth-pruned submodel and binary matrix to obtain the structured-pruned submodel. 
The structured pruning strategy significantly reduces the capacity of model, saves storage and computation resources, and accelerates directly on resource-limited clients. 

\begin{algorithm}[t]
  \small
  \caption{${Co\text{-}S}^2{P}$ client-training}
  \label{algo:client_training}
  \SetKwComment{Comment}{// }{}
  \SetKwInput{Input}{Input}
  \Input{datasets $\mathcal{D}_n$, current round $q$, weights $w_{0,n}$ of depth-pruned submodel, rounds $\hat{Q}$, local epochs $\hat{E}$ and learning rate $\hat{\eta}$ for training masks, local epochs $E$ and learning rate $\eta$ for training submodels}
  \textbf{Initialize:} the important mask $I_{n}$\;

  \If{$q < \hat{Q}$}{
    \For{epoch $e=1$ \KwTo $\hat{E}$}{
      \For{segment $i \in w_{n}$}{
        $M_n^{l_i} \sim \mathrm{Bern}\!\Big(\frac{1}{1+e^{-I_n^{l_i}}}\Big)$\;
        Obtain $\hat{M}_{n}^{l_i}$ by extending $M_{n}^{l_i}$\;
        $\widetilde w_{0,n}^{l_i} = w_{0,n}^{l_i} \odot \hat{M}_{n}^{l_i}$ \Comment*[r]{Freeze $w_{0,n}$}
      }
      $I_{n} = I_{n} - \hat{\eta}\,\nabla_{I_{n}} \mathcal L_{n,\text{mask}}(\widetilde w_{0,n} ; \mathcal{D}_n)$\;
    }
  }
  $\hat{w}_{0,n} \leftarrow w_{0,n} \odot \hat{M}_{n}$\;
  \For{epoch $e=1$ \KwTo $E$}{
    $\hat{w}_{e,n} = \hat{w}_{e-1,n} - \eta\, \nabla_{w_{e-1,n}} \mathcal L_{n,\text{weight}}(\hat{w}_{e-1,n} ; \mathcal{D}_n)$\;
  }
\end{algorithm}

\subsection{Self-Distillation based Knowledge Transfer}\label{CBKT}

\begin{figure}[t]
  \centering
  \includegraphics[width=0.5\textwidth]{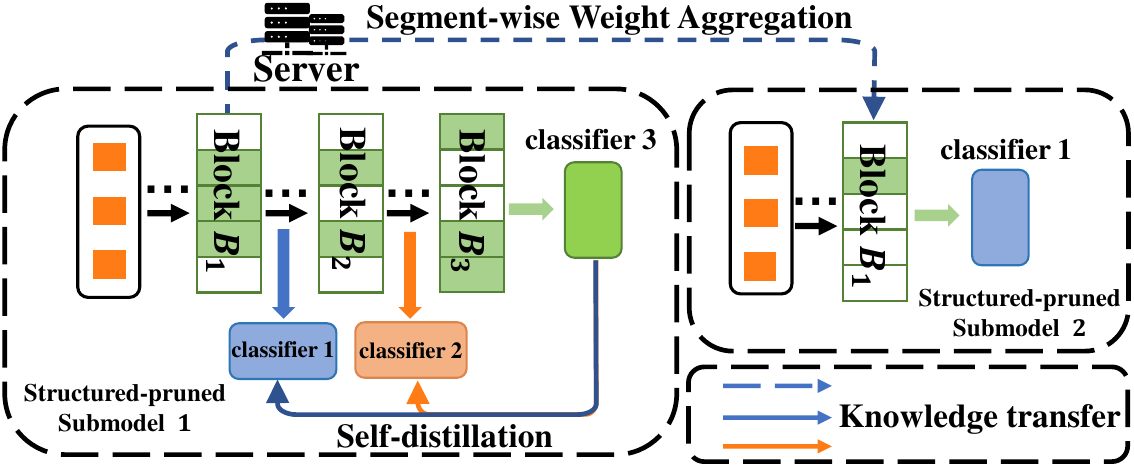}
   \caption{Cross-block knowledge transfer. In structured-pruned submodel 1, the high-level knowledge transfers from block $B_3$ to blocks $B_1$ and $B_2$ through self-distillation. Through the segment-wise weight aggregation strategy, the high-level knowledge is further transferred in blocks at the same location.}
  \label{fig:corss_block}
\end{figure}

To mitigate the knowledge loss problem in resource-limited clients, $Co\text{-}S^2P$ introduces lightweight self-distillation mechanism~\cite{zhang2019your} to implement cross-block knowledge transfer, as shown in Fig.\ref{fig:corss_block}.
For the structured-pruned submodel, we place multiple classifiers at specific depths $h$, where $h \in \{h^{frozen} + L\times {R_i^{depth}} | {R_i^{depth}}\leq{R_n^{depth}}, 1 \leq i \leq N\}$, and $h^{frozen}$ is the number of the frozen blocks and $L$ is the number of blocks in the global model. We use self-distillation by treating the deepest classifier as the teacher and other classifiers as the students. The loss function for the width pruning is as follows:
\begin{equation}
\begin{aligned}
\mathcal L_{n, weight} = \sum_{i=1}^{c_n} ((1-\lambda_2)\mathcal L_{CE}(\hat{y}_i, y) + \lambda_2 \mathcal L_{KL}(\hat{y}_i, \hat{y}_{c_n})),
\end{aligned}
\end{equation}
where $c_n$ is the number of classifiers in client $n$, $\mathcal L_{CE}(\hat{y}_i, y)$ represents the cross entropy loss of the $i$-th classifier for target class $y$. $\mathcal L_{KL}(\hat{y}_i, \hat{y}_{c_n})=sum(\sigma(\hat{y}_i/t)\ln{\frac{\sigma(\hat{y}_i/t)}{\sigma(\hat{y}_{c_n}/t)}})t^2$, where temperature $t$ is a hyperparameter that controls the information capacity of data distributions provided by the teacher submodel, $\lambda_2$ indicates the balance between the cross entropy loss and self-distillation. 

\textcolor{\attcolor}{
The self-distillation mechanism enables the shallow blocks to be equipped with high-level knowledge without extra overhead, because the student model shares the same model backbone as the teacher model and is trained by reusing its forward propagated activations and back-propagated gradients.
When paired with the designed data distribution‑aware structured pruning, the mechanism enables personalized submodels to adapt to client-specific resource and heterogeneous data as demonstrated in our experiments (Sec.\ref{exp:ablation}).}
The local training process of each client is outlined in Algorithm \ref{algo:client_training}.

\vspace{-0.5em}
\subsection{Semi-Asynchronous Aggregation Strategy}\label{SAS}
\textcolor{\attcolor}{Due to the stragglers, we design a pruning-based semi-asynchronous aggregation strategy with  minimum ratio $\mu$ of received clients and waiting interval $T_{clk}$ 
to balance the performance of the collaborative training and the resource utilization of all clients, which is shown in Algorithm~\ref{algo:server_aggregation}.
After the server receives $\mu N$ client updates, it turns on the clock timing and waits for $T_{clk}$ seconds before aggregation.  }

\textcolor{\attcolor}{Considering that some updates of varying submodels across clients are stale and thus deviate significantly from the latest global model, we compute the segment importance score across clients training segment $i$
as follows:}
\begin{equation}
\label{eq:importantcomputing}
\gamma_{n}^i = \frac{\|\Delta_{q,n}^i\|_1}{\|w_q^i - w_{q-\tau_{q,n}}^i\|_1 + size(w_q^i)},
\end{equation}
where $\Delta_{q,n}^i$ is the gradients of the client $n$ on segment $i$ in $q$-th round that indicates the extent of updating.
$w_{q-\tau_{q,n}}^i$ represents the weights of the global model on segment $i$ in the $(q-\tau_{q,n})$-th round, $\tau_{q,n}$ is the delay rounds that the client $n$ has not taken part in global aggregation. $\|w_q^i - w_{q-\tau_{q,n}}^i\|_1$ indicates the difference of the received model by the client $n$ and the latest model.

\begin{algorithm}[t]
  \small
  \caption{Semi-Async. Weight Aggregation}	
  \label{algo:server_aggregation}
  \SetKwProg{Function}{Function}{}{}
  \SetKwInput{Input}{Input}
  \Input{initial weight $w^0$, aggregation minimal ratio $\mu$, waiting interval $T_{clk}$, total communication round $Q$}
  \For{$q=1$ {\bfseries to} $Q$}
  {
    $C_q \leftarrow$ available Clients  \\
    \For{$C_{n,q} \in C_q$ (all workers in parallel)}
    {
        ${Co\text{-}S}^2{P}$ client-training($w_{n,q}$, $q$)  \\
    }
    $cnt = 0$ \\
    \While{$cnt < \mu N$}
    {
        $\Delta_{q,n}, M_{q,n}, \tau_{q,n} \leftarrow$ update from client $n$ \\
        \For{Segment $i \in \Delta_{n}$}
        {
            compute $\gamma_{n}^i$ via Eq.\ref{eq:importantcomputing}  \\
        }
        $cnt = cnt + 1$  \\
    }
    $t_{clk} \leftarrow$ Current time, $T_{timeout} = t_{clk} + T_{clk}$  \\
    \While{$t_{clk} < T_{timeout}$}
    {
        $\Delta_{q,n}, M_{q,n}, \tau_{q,n} \leftarrow$ update from client $n$ \\
        \For{Segment $i \in \Delta_{n}$} 
        {
            compute $\gamma_{n}^i$ via Eq.\ref{eq:importantcomputing}  \\
        }
        $t_{clk} \leftarrow$ Current time   \\
    }
    \For{segment $i \in w$} 
    {
        $N_q^i = \{n: M_{n,q}^i=1\}$  \\
        $ w_{q+1}^i = w_q^i - \eta \sum_{n \in N_q^i} \frac{\gamma_n^i}{\sum_{n\in N_q^i}\gamma_{n}^i} \Delta_{n}^i$   \\
    }
  }
\end{algorithm}

Finally, the server aggregates the gradients and updates segment $i$ by normalizing the segment important scores on segment $i$ as follows:
\begin{equation}
w_{q+1}^i = w_q^i - \eta \sum_{n \in N_q^i} \frac{\gamma_n^i}{\sum_{n\in N_q^i}\gamma_{n}^i} \Delta_{n}^i,
\end{equation}
where $N_q^i=\{n: M_{n,q}^i=1\}$ represents the set of clients training segment $i$ in the $q$-th round, $\eta$ is the learning rate for training the masked submodel. 
It is worth noting that the aggregation strategy also transfers the high-level knowledge from the deep blocks to shallow blocks, as shown in Fig.\ref{fig:corss_block}.


\section{Convergence Analysis}\label{con_ana}
In this section, we show the convergence rate of our proposed pruning-based semi-asynchronous collaborative learning framework 
${Co\text{-}S}^2{P}$. Firstly, we give a crucial definition for theoretical analysis:
\begin{definition}\label{Number of non-participating rounds}
(Maximum delay).
 \textit{We define the delay as the client $n$ has not taken part in the global aggregation for $\tau_{q,n}$ rounds, so we can get:}
\begin{equation}\label{Number of non-trained rounds}
\tau_{q}=\max\limits_{n}\tau_{q,n},\quad n\in N.
\end{equation}
\end{definition}

Then, we give assumptions for ease of convergence analysis:
\begin{assumption}\label{Lipschitzian Condition}
	($L$-smooth). 
        \textit{Every function $F_n(\cdot)$ is $L$-smooth for all $i\in [N],w,v\in R^d$ }
        \begin{equation}
             \| \nabla F_{n}(w) - \nabla F_{n}(v)\| \leq L\|w - v\|.
        \end{equation}
\end{assumption}

\begin{assumption}\label{Bounded variance}
	(Bounded variance). \textit{There exists $\sigma>0$:}
\begin{equation}\label{S-V}
    	\mathbb{E}_{\xi_{i}\sim D_{n}}\|\nabla F_{n}(w;\xi_{i}) - \nabla F_{n}(w)\|^{2} \leq \sigma^{2},
\end{equation}
 \textit{$\sigma>0$ bounds the variance of stochastic gradient.}
\end{assumption}

\begin{assumption}\label{Bounded Non-IID level}
	(Bounded data heterogeneity level). \textit{There exists $\delta>0$:}
\begin{equation}
    	\|\nabla F_{n}(w_q) - \nabla F(w_q)\|^{2} \leq \delta^{2},
\end{equation}
 \textit{$\delta>0$ bounds the effect of heterogeneous data.}
\end{assumption}


\begin{assumption}\label{Bounded gradient}
    (Bounded gradient).
    \textit{In algorithm 2, the expected squared norm of stochastic gradients is bounded uniformly, for constant $G>0$ and $\forall n,q,t$:}
    \begin{equation}
    \mathbb{E}\|\nabla F_n(w_{q,n,t},\xi_{q,n,t})\|^2\leq G.
    \end{equation}
\end{assumption}

\begin{remark}
\textcolor{\attcolor}{The above assumptions are commonly used in previous works. Assumption~\ref{Lipschitzian Condition} imposes the levels of Lipchitz smoothness, which is a standard assumption in distributed learning~\cite{huang2024distributed}. Assumption~\ref{Bounded variance} shows the stochastic estimators are unbiased with bounded variances, which is widely used in stochastic optimization~\cite{cho2023convergence}. Assumption~\ref{Bounded Non-IID level} quantifies the heterogeneity among the local datasets, which is common in traditional federated learning~\cite{li2019convergence,yuan2022convergence}. Assumption~\ref{Bounded gradient} gives an upper bound on the norm of the gradients, which is also adopted by other studies~\cite{zhou2023every}.}
\end{remark}
With these definitions and assumptions, we can bound the deviation of the average squared gradient norm for the convergence analysis of ${Co\text{-}S}^2{P}$.
\begin{theorem}\label{theorem 2}
	Let all assumptions hold. Suppose that the step size $\eta$ satisfies the following relationships:
\begin{equation*}\label{}
   \left\{\begin{array}{r c l}
    8\eta^2L^2E^2\leq\frac{1}{2}\Rightarrow \eta\leq\frac{1}{4LE} \nonumber\\
     \frac{L}{2}\eta^2E^2-\frac{E\eta}{2}<0\Rightarrow \eta<\frac{1}{LE}\nonumber
    \end{array}\right..
\end{equation*}

Therefore, the step size $\eta$ is defined as:
   $ 0\leq \eta \leq \frac{1}{4LE}.$
   
   Then, for all $Q\geqslant 1$, we have :
\begin{small}
\begin{align*} 	
\frac{1}{Q}\sum_{q=1}^Q\mathbb{E}\|\nabla F(w_q)\|^2& \leq \frac{2\mathbb{E}[F(w_{1})]}{E\eta Q}\\
&+4L\eta^2E^2\frac{N|K|}{N^*}G(\tau \frac{N|K|}{N^*}+16)\\
&+16L\eta^2E\frac{N|K|}{N^*}\sigma^2+64L\eta^2E^2\frac{N|K|}{N^*}\delta^2,
\end{align*} 
\end{small}
where $\frac{1}{Q}\sum_{q=1}^Q(\tau_q)^2=\tau$, $|K|$ is the number of segment, $N^*=\min_{q,i}|N_q^i|$, means the minimum number of submodels training the corresponding segment $i$ in all rounds.
\end{theorem}
\begin{remark}
\textcolor{\attcolor}{Theorem 1 bounds the averaged gradient across all communication rounds. The upper bound includes four parts. The first term $\frac{2\mathbb{E}[F(w_{1})]}{E\eta Q}$ represents the initial function value's contribution to the convergence.
This term diminishes as the number of communication rounds $Q$, the number of local steps $E$, and the learning rate $\eta$ increase. 
The second term $4L\eta^2E^2\frac{N|K|}{N^*}G(\tau \frac{N|K|}{N^*}+16)$ captures the error accumulated from the boundedness of stochastic gradients $G$. 
The third term $16L\eta^2E\frac{N|K|}{N^*}\sigma^2$ is directly related to the variance of the stochastic gradients $\sigma^2$, quantifying the noise introduced by using mini-batch or sampled gradients.
Finally, the fourth term $64L\eta^2E^2\frac{N|K|}{N^*}\delta^2$ accounts for the impact of data heterogeneity $\delta^2$ across different nodes. This term highlights how the non-IID nature of data in a distributed setting contributes to the overall convergence bound.
The last three term contribute to an error floor independent of $Q$ that does not vanish with increasing $Q$.
}    
\end{remark}

Theorem~\ref{theorem 2} shows the convergence of the semi-asynchronous aggregation mechanism in ${Co\text{-}S}^2{P}$ with the upper bound on the average gradient of all segments. \textcolor{\attcolor}{We denote the number of clients $N$, the local update epoch $E$ and the number of segments $|K|$ as constants. We can gain the following remarks.}

\begin{remark}
\textbf{Impact of the maximum delay $\tau_q$.} As we defined, $\tau_q=\max\limits_{n}\tau_{q,n}$ is the maximum delay between all clients until round $q$. The result indicates that the larger $\tau_q$ is, the worse the convergence rate is. Taking into account the local training of the submodel, the larger $\tau_q$ means that the initial local submodels are more biased towards the latest global model. And a smaller $\tau_q$ is beneficial to the convergence but disturbed by stragglers. Therefore, our semi-asynchronous aggregation balances the convergence rate and the resource utilization.

\end{remark}

\begin{remark}
\textcolor{\attcolor}{\textbf{Impact of data heterogeneity $\delta$.} As we described, the data distribution is always heterogeneous in a real-world setting. Theorem 1 demonstrates that data heterogeneity is a key factor in affecting convergence. The larger $\delta$ denotes the higher data heterogeneity, which can slow the convergence rate. When degenerated to the iid s.t. $\delta=0$, this term becomes zero, which is faster than the existing convergence rate.}
\end{remark}

\textit{Proof.} Let us start the proof of the global model generated by semi-asynchronous aggregation strategy from $L$-Lipschitz Condition:
\begin{small}
\begin{align*} 	
    \mathbb{E}[F(w_{q+1})]-\mathbb{E}[F(w_{q})]&\leq \underbrace{\mathbb{E}[\langle\nabla F(w_{q}),w_{q+1}-w_{q}\rangle]}_{U_1}\\
    &+\underbrace{\frac{L}{2}\mathbb{E}\|w_{q+1}-w_{q}\|^2}_{U_2}   
\end{align*} 
\end{small}
bound $U_1$:
\begin{small}
\begin{align*} 	
    &\mathbb{E}[\langle\nabla F(w_{q}),w_{q+1}-w_{q}\rangle]\\
    &=\sum_{i\in S_q}\mathbb{E}[\langle\nabla F^i(w_{q}),w_{q+1}^i-w_{q}^i\rangle]\\
    &+\sum_{i\in K-S_q}\mathbb{E}[\langle\nabla F^i(w_{q}),w_{q+1}^i-w_{q}^i\rangle]\\
    &=\sum_{i\in S_q}\mathbb{E}[\langle\nabla F^i(w_{q}),w_{q+1}^i-w_{q}^i\rangle]+\sum_{i\in K-S_q}\mathbb{E}[\langle\nabla F^i(w_{q}),\mathbf{0}\rangle]\\
    &=\sum_{i\in S_q}\mathbb{E}[\langle\nabla F^i(w_{q}),w_{q+1}^i-w_{q}^i\rangle]\\
     &=\sum_{i\in S_q}\mathbb{E}[\langle\nabla F^i(w_{q}),-\eta (\sum\limits_{n\in N_{q}^i}\Tilde{\gamma}_{q,n}^i\Delta_{q,n}^i)\rangle]\\  
     &=\sum_{i\in S_q}\mathbb{E}[\langle\nabla F^i(w_{q}),-\eta (\sum\limits_{n\in N_{q}^i}(\Tilde{\gamma}_{q,n}^i\frac{(w_{q-\tau_q,n,0}^i-w_{q-\tau_q,n,E}^i)}{\eta}))\rangle]\\ 
     &=\sum_{i\in S_q}\mathbb{E}[\langle\nabla F^i(w_{q}),-\eta (\sum\limits_{n\in N_{q}^i}\sum_{e=1}^{E}\Tilde{\gamma}_{q,n}^i\nabla F_n^i(w_{q-\tau_q,n,e-1},\xi_{n,e-1}))\rangle]\\   
     &=-E\eta\sum_{i\in S_q}\mathbb{E}[\langle\nabla F^i(w_{q}),\sum\limits_{n\in N_{q}^i}\frac{1}{E}\sum_{e=1}^{E}\Tilde{\gamma}_{q,n}^i\nabla F_n^i(w_{q-\tau_q,n,e-1},\xi_{n,e-1})\rangle]\\
     &\stackrel{(a)}{=}-E\eta\sum_{i\in S_q}[\frac{1}{2}\mathbb{E}\|\nabla F^i(w_{q})\|^2\\
     &+\frac{1}{2}\mathbb{E}\|\sum\limits_{n\in N_{q}^i}\frac{1}{E}\sum_{e=1}^{E}\Tilde{\gamma}_{q,n}^i\nabla F_n^i(w_{q-\tau_q,n,e-1},\xi_{n,e-1})\|^2\\
     &-\frac{1}{2}\mathbb{E}\|\nabla F^i(w_{q})-\sum\limits_{n\in N_{q}^i}\frac{1}{E}\sum_{e=1}^{E}\Tilde{\gamma}_{q,n}^i\nabla F_n^i(w_{q-\tau_q,n,e-1},\xi_{n,e-1})\|^2]\\
     &=-\frac{E\eta}{2}\sum_{i\in S_q}\mathbb{E}\|\nabla F^i(w_{q})\|^2\\
     &-\frac{E\eta}{2}\sum_{i\in S_q}\mathbb{E}\|\sum\limits_{n\in N_{q}^i}\frac{1}{E}\sum_{e=1}^{E}\Tilde{\gamma}_{q,n}^i\nabla F_n^i(w_{q-\tau_q,n,e-1},\xi_{n,e-1})\|^2\\
     &+\frac{E\eta}{2}\underbrace{\sum_{i\in S_q}\mathbb{E}\|\nabla F^i(w_{q})-\sum\limits_{n\in N_{q}^i}\frac{1}{E}\sum_{e=1}^{E}\Tilde{\gamma}_{q,n}^i\nabla F_n^i(w_{q-\tau_q,n,e-1},\xi_{n,e-1})\|^2}_{U_3}
\end{align*} 
\end{small}
where (a) is from the equation $\langle a,b\rangle=\frac{1}{2}(\|a\|^2+\|b\|^2-\|a-b\|^2)$ with $a=\nabla F^i(w_{q})$ and $b=\sum\limits_{n\in N_{q}^i}\frac{1}{E}\sum_{e=1}^{E}\Tilde{\gamma}_{q,n}^i$ $\nabla F_n^i(w_{q-\tau_q,n,e-1},\xi_{n,e-1})$.

To bound $U_3$:
\begin{small}
\begin{align*}
&\sum_{i\in S_q}\mathbb{E}\|\nabla F^i(w_{q})-\sum\limits_{n\in N_{q}^i}\frac{1}{E}\sum_{e=1}^{E}\Tilde{\gamma}_{q,n}^i\nabla F_n^i(w_{q-\tau_q,n,e-1},\xi_{n,e-1})\|^2\\
&=\sum_{i\in S_q}\mathbb{E}\|\sum\limits_{n\in N_{q}^i}\frac{1}{E}\sum_{e=1}^{E}\Tilde{\gamma}_{q,n}^i(\nabla F_n^i(w_{q})-\nabla F_n^i(w_{q-\tau_q,n,e-1},\xi_{n,e-1}))\|^2\\ 
&=\sum_{i\in S_q}\mathbb{E}\|\sum\limits_{n\in N_{q}^i}\frac{1}{E}\sum_{e=1}^{E}\Tilde{\gamma}_{q,n}^i(\nabla F_n^i(w_{q})-\nabla F_n^i(w_{q-\tau_q,n,e-1})\\
&+\nabla F_n^i(w_{q-\tau_q,n,e-1})-\nabla F_n^i(w_{q-\tau_q,n,e-1},\xi_{n,e-1}))\|^2\\
&\leq 2\sum_{i\in S_q}\mathbb{E}\|\sum\limits_{n\in N_{q}^i}\frac{1}{E}\sum_{e=1}^{E}\Tilde{\gamma}_{q,n}^i(\nabla F_n^i(w_{q})-\nabla F_n^i(w_{q-\tau_q,n,e-1}))\|^2\\
&+2\sum_{i\in S_q}\mathbb{E}\|\sum\limits_{n\in N_{q}^i}\frac{1}{E}\sum_{e=1}^{E}\Tilde{\gamma}_{q,n}^i(\nabla F_n^i(w_{q-\tau_q,n,e-1})\\
&-\nabla F_n^i(w_{q-\tau_q,n,e-1},\xi_{n,e-1}))\|^2\\
&\leq 2\sum_{i\in S_q}\sum\limits_{n\in N_{q}^i}\frac{1}{E}\sum_{e=1}^{E}\Tilde{\gamma}_{q,n}^i\mathbb{E}\|\nabla F_n(w_{q})-\nabla F_n(w_{q-\tau_q,n,e-1})\|^2\\
&+2\sum_{i\in S_q}\sum\limits_{n\in N_{q}^i}\frac{1}{E}\sum_{e=1}^{E}\Tilde{\gamma}_{q,n}^i\\
&*\mathbb{E}\|\nabla F_n(w_{q-\tau_q,n,e-1})-\nabla F_n(w_{q-\tau_q,n,e-1},\xi_{n,e-1})\|^2\\
&\stackrel{(a)}{\leq} 2L\sum_{i\in S_q}\sum\limits_{n\in N_{q}^i}\frac{1}{E}\sum_{e=1}^{E}\Tilde{\gamma}_{q,n}^i\mathbb{E}\|w_{q}-w_{q-\tau_q,n,e-1}\|^2+2\sigma^2\\
&\leq 4L\sum_{i\in S_q}\sum\limits_{n\in N_{q}^i}\Tilde{\gamma}_{q,n}^i\frac{1}{E}\sum_{e=1}^{E}\underbrace{\mathbb{E}\|w_{q}-w_{q-\tau_q}\|^2}_{U_4}\\
&+4L\sum_{i\in S_q}\sum\limits_{n\in N_{q}^i}\Tilde{\gamma}_{q,n}^i\underbrace{\frac{1}{E}\sum_{e=1}^{E}\mathbb{E}\|w_{q-\tau_q}-w_{q-\tau_q,n,e-1}\|^2}_{U_5}+2\sigma^2
\end{align*}
\end{small}
where (a) is from the Assumption~\ref{Bounded variance}.

To bound $U_4$:
\begin{small}
\begin{align*} 	
\mathbb{E}\|w_{q}-w_{q-\tau_q}\|^2&=\sum_{i\in S_q}\mathbb{E}\|w_{q}^i-w_{q-\tau_q}^i\|^2+\sum_{i\in K-S_q}\mathbb{E}\|w_{q}^i-w_{q-\tau_q}^i\|^2\\
&\stackrel{(a)}{\leq} \sum_{i\in S_q} \tau_{q}\sum_{l=0}^{\tau_q-1}\mathbb{E}\|w_{q-l}^i-w_{q-(l+1)}^i\|^2\\
&+\sum_{i\in K-S_q} \tau_{q}\sum_{l=0}^{\tau_q-1}\mathbb{E}\|w_{q-l}^i-w_{q-(l+1)}^i\|^2\\
&\leq \tau_{q}\sum_{l=0}^{\tau_q-1} \underbrace{\sum_{i\in S_q}\mathbb{E}\|w_{q-l}^i-w_{q-(l+1)}^i\|^2}_{U_6}\\
&+\tau_{q}\sum_{l=0}^{\tau_q-1}\underbrace{\sum_{i\in K-S_q} \mathbb{E}\|w_{q-l}^i-w_{q-(l+1)}^i\|^2}_{U_7}
\end{align*} 
\end{small}
where (a) is from the definition of $\tau_q$. 

Here, we define the normalized gamma as $\Tilde{\gamma}_{q,n}^i= \frac{\gamma_n^i}{\sum_{n\in N_q^i}\gamma_{n}^i}.$

To bound $U_6$:
 \begin{small}
\begin{align*} 	
&\sum_{i\in S_q}\mathbb{E}\|w_{q-l}^i-w_{q-(l+1)}^i\|^2\\
&=\sum_{i\in S_q}\mathbb{E}\|-\eta (\sum\limits_{n\in N_{q-(l+1)}^i}\Tilde{\gamma}_{q-(l+1),n}^i\Delta_{q-(l+1),n}^i)\|^2\\
&=\eta^2\sum_{i\in S_q}\mathbb{E}\|\sum\limits_{n\in N_{q-(l+1)}^i}\Tilde{\gamma}_{q-(l+1),n}^i\Delta_{q-(l+1),n}^i)\|^2\\
&=\eta^2\sum_{i\in S_q}\mathbb{E}\|\sum\limits_{n\in N_{q-(l+1)}^i}\sum_{e=1}^{E}\Tilde{\gamma}_{q-(l+1),n}^i\\
&*\nabla F_n^i(w_{q-(l+1)-\tau_{q-(l+1)},n,e-1},\xi_{n,e-1})\|^2\\
&\leq \eta^2E\sum_{i\in S_q}\sum\limits_{n\in N_{q-(l+1)}^i}\Tilde{\gamma}_{q-(l+1),n}^i\\
&*\sum_{e=1}^{E}\mathbb{E}\|\nabla F_n^i(w_{q-(l+1)-\tau_{q-(l+1)},n,e-1},\xi_{n,e-1})\|^2\\
&\stackrel{(a)}{\leq} \eta^2E\sum_{i\in S_q}\sum\limits_{n\in N_{q-(l+1)}^i}\frac{1}{N_{q-(l+1)}^i}EG\\
&\leq \eta^2E^2 \frac{N}{N^*}|S_q|G 
\end{align*} 
\end{small}
where (a) is from Assumption~\ref{Bounded gradient}.

To bound $U_7$:
\begin{small}
\begin{align*} 	
&\sum_{i\in K-S_q}\mathbb{E}\|w_{q-l}^i-w_{q-(l+1)}^i\|^2\\
&=\sum_{i\in C_{q-(l+1)}}\mathbb{E}\|w_{q-l}^i-w_{q-(l+1)}^i\|^2\\
&+\sum_{i\in K-S_q-C_{q-(l+1)}}\mathbb{E}\|w_{q-l}^i-w_{q-(l+1)}^i\|^2\\
&=\sum_{i\in C_{q-(l+1)}}\mathbb{E} | -\eta (\sum\limits_{n\in N_{q-(l+1)}^i}\Tilde{\gamma}_{q-(l+1),n}^i\Delta_{q-(l+1),n}^i)\|^2+\textbf{0}\\
&=\eta^2\sum_{i\in C_{q-(l+1)}}\mathbb{E}\|\sum\limits_{n\in N_{q-(l+1)}^i}\sum_{e=1}^{E}\Tilde{\gamma}_{q-(l+1),n}^i\\
&*\nabla F_n^i(w_{q-(l+1)-\tau_{q-(l+1)},n,e-1},\xi_{n,e-1})\|^2\\
&\leq\eta^2E\sum_{i\in C_{q-(l+1)}}\sum\limits_{n\in N_{q-(l+1)}^i}\Tilde{\gamma}_{q-(l+1),n}^i\\
&*\sum_{e=1}^{E}\mathbb{E}\|\nabla F_n^i(w_{q-(l+1)-\tau_{q-(l+1)},n,e-1},\xi_{n,e-1})\|^2\\
&\stackrel{(a)}{\leq} \eta^2 E^2\frac{N}{N^*}|C_{q-(l+1)}|G
\end{align*} 
\end{small}
where $C_{q-(l+1)}$ is the region in $K-S_q$ that has been trained in round $q-(l+1)$. (a)) is from Assumption~\ref{Bounded gradient}.

Plugging $U_6, U_7$ into $\mathbb{E}\|w_{q}-w_{q-\tau_q}\|^2$ and using $|S_q|+|C_{q-(l+1)}|\leq |K|$, we can gain the bound of $U_4$:
\begin{equation}
    \mathbb{E}\|w_{q}-w_{q-\tau_q}\|^2\leq (\tau_q)^2\eta^2E^2\frac{N}{N^*}|K|G
\end{equation}

To bound $U_5$:
\begin{small}
\begin{align*} 	
&\frac{1}{E}\sum_{e=1}^E\mathbb{E}\|w_{q-\tau_q,n,e-1}-w_{q-\tau_q}\|^2\\
&=\frac{1}{E}\sum_{e=1}^E\mathbb{E}\|w_{q-\tau_q,n,e-1}-w_{q-\tau_q,n,0}\|^2\\
&=\frac{1}{E}\sum_{e=1}^E\mathbb{E}\|\sum_{j=0}^{e-2}-\eta\nabla F_n(w_{q-\tau_q,n,j},\xi_{n,j})\odot M_{q-\tau_q,n}\|^2\\
&=\frac{\eta^2}{E}\sum_{e=1}^E\mathbb{E}\|\sum_{j=0}^{e-2}(\nabla F_n(w_{q-\tau_q,n,j},\xi_{n,j})-\nabla F_n(w_{q-\tau_q,n,j})\\
&+\nabla F_n(w_{q-\tau_q,n,j}))\odot M_{q-\tau_q,n}\|^2\\
&\leq \frac{2\eta^2}{E}\sum_{e=1}^E\mathbb{E}\|\sum_{j=0}^{e-2}(\nabla F_n(w_{q-\tau_q,n,j},\xi_{n,j})\\&\quad\quad\quad-\nabla F_n(w_{q-\tau_q,n,j}))\odot M_{q-\tau_q,n}\|^2\\
&+\frac{2\eta^2}{E}\sum_{e=1}^E\mathbb{E}\|\sum_{j=0}^{e-2}\nabla F_n(w_{q-\tau_q,n,j})\odot M_{q-\tau_q,n}\|^2\\
&\stackrel{(a)}{\leq} \frac{2\eta^2}{E}\sum_{e=1}^E(e-1)\sigma^2+\frac{2\eta^2}{E}\sum_{e=1}^E\mathbb{E}\|\sum_{j=0}^{e-2}(\nabla F_n(w_{q-\tau_q,n,j})\\
&- \nabla F_n(w_{q-\tau_q})+\nabla F_n(w_{q-\tau_q}))\odot M_{q-\tau_q,n}\|^2\\
&\leq 2\eta^2 E\sigma^2\\
&+\frac{4\eta^2}{E}\sum_{e=1}^E (e-1)\sum_{j=0}^{e-2}\mathbb{E}\|(\nabla F_n(w_{q-\tau_q,n,j})\\&\quad\quad\quad-\nabla F_n(w_{q-\tau_q}))\odot M_{q-\tau_q,n}\|^2\\
&+\frac{4\eta^2}{E}\sum_{e=1}^E (e-1)\sum_{j=0}^{e-2}\mathbb{E}\|\nabla F_n(w_{q-\tau_q})\odot M_{q-\tau_q,n}\|^2\\
&\stackrel{(b)}{\leq} 2\eta^2 E\sigma^2+\frac{4\eta^2 L^2}{E}\sum_{e=1}^E (e-1)\sum_{j=0}^{e-2}\mathbb{E}\|w_{q-\tau_q,n,j}-w_{q-\tau_q}\|^2\\
&+4\eta^2 E^2\mathbb{E}\|(\nabla F_n(w_{q-\tau_q})-\nabla F(w_{q-\tau_q})+\nabla F(w_{q-\tau_q}))\odot M_{q-\tau_q,n}\|^2\\
&\leq 2\eta^2 E\sigma^2+4\eta^2 L^2E(E-1)\frac{1}{E}\sum_{e=1}^E\mathbb{E}\|w_{q-\tau_q,n,t-1}-w_{q-\tau_q}\|^2\\
&+8\eta^2 E^2\mathbb{E}\|(\nabla F_n(w_{q-\tau_q})-\nabla F(w_{q-\tau_q}))\odot M_{q-\tau_q,n}\|^2\\
&+ 8\eta^2 E^2\mathbb{E}\|\nabla F(w_{q-\tau_q})\odot M_{q-\tau_q,n}\|^2\\
&\stackrel{(c)}{\leq} 2\eta^2 E\sigma^2+4\eta^2 L^2E^2\frac{1}{E}\sum_{e=1}^E\mathbb{E}\|w_{q-\tau_q}-w_{q-\tau_q,n,e-1}\|^2\\
&+8\eta^2 E^2 \delta^2+8\eta^2 E^2\mathbb{E}\|\nabla F(w_{q-\tau_q})\odot M_{q-\tau_q,n}\|^2
\end{align*} 
\end{small}
where (a) is from Assumption~\ref{Bounded variance}, (b) is from Assumption~\ref{Lipschitzian Condition} and (c) is from Assumption~\ref{Bounded Non-IID level}.

Let\quad  $4\eta^2L^2E^2\leq\frac{1}{2}\Rightarrow \eta\leq\frac{1}{2LE}$, we can get:
\begin{small}
\begin{flalign*} 	
&\frac{1}{E}\sum_{e=1}^E\mathbb{E}\|w_{q-\tau_q,n,t-1}-w_{q-\tau_q}]\|^2\\
&\stackrel{(a)}{\leq} 4\eta^2 E\sigma^2+16\eta^2 E^2 \delta^2+16\eta^2 E^2\mathbb{E}\|\nabla F(w_{q-\tau_q})\odot M_{q-\tau_q,n}\|^2\\
&\leq 4\eta^2 E\sigma^2+16\eta^2 E^2 \delta^2+16\eta^2 E^2\sum_{i\in S_{q-\tau_q}}\mathbb{E}\|\nabla F^i(w_{q-\tau_q})\|^2\\
&\stackrel{(b)}{\leq} 4\eta^2 E\sigma^2+16\eta^2 E^2 \delta^2+16\eta^2 E^2G
\end{flalign*} 
\end{small}
where (a) is due to the fact that $\eta\leq\frac{1}{2LE}$ and (b) is from Assumption~\ref{Bounded gradient}.

Plugging $U_4,U_5$ into $U_3$, we can gain:
\begin{small}
\begin{align*}
& \sum_{i\in S_q}\mathbb{E}\|\nabla F^i(w_{q})-\sum\limits_{n\in N_{q}^i}\frac{1}{E}\sum_{e=1}^{E}\Tilde{\gamma}_{q,n}^i\nabla F_n^i(w_{q-\tau_q,n,e-1},\xi_{n,e-1})\|^2\\
&\leq 4L\sum_{i\in S_q}\sum\limits_{n\in N_{q}^i}\Tilde{\gamma}_{q,n}^i\frac{1}{E}\sum_{e=1}^{E} 2(\tau_q)^2\eta^2E^2NG\\
&+4L\sum_{i\in S_q}\sum\limits_{n\in N_{q}^i}\Tilde{\gamma}_{q,n}^i(4\eta^2 E\sigma^2+16\eta^2 E^2 \delta^2+16\eta^2 E^2G)+2\sigma^2\\
&\leq 4L(\tau_q)^2\eta^2E^2(\frac{N}{N^*})^2|K|^2G+16L\eta^2E\frac{N}{N^*}|K|\sigma^2\\
&+64L\eta^2E^2\frac{N}{N^*}|K|\delta^2+64L\eta^2E^2\frac{N}{N^*}|K|G 
\end{align*}
\end{small}
To bound $U_2$:
\begin{small}
\begin{align*}
&\frac{L}{2}\mathbb{E}\|w_{q+1}-w_{q}\|^2\\
&=\frac{L}{2}\sum_{i\in S_q}\mathbb{E}\|w_{q+1}^i-w_{q}^i\|^2 + \frac{L}{2}\sum_{i\in K-S_q}\mathbb{E}\|w_{q+1}^i-w_{q}^i\|^2\\
&=\frac{L}{2}\sum_{i\in S_q}\mathbb{E}\|w_{q+1}^i-w_q^i\|^2\\
&=\frac{L}{2}\sum_{i\in S_q}\mathbb{E}\|\sum_{n\in N_q^i}\Tilde{\gamma}_{q,n}^i(w_{q,n,0}-w_{q,n,E})^i\|^2\\
&=\frac{L\eta^2E^2}{2}\sum_{i\in S_q}\mathbb{E}\| \sum\limits_{n\in N_{q}^i}\frac{1}{E}\sum_{e=1}^{E}\Tilde{\gamma}_{q,n}^i\nabla F_n^i(w_{q-\tau_q,n,e-1},\xi_{n,e-1})\|^2
\end{align*}
\end{small}
So, we can get:
\begin{small}
\begin{align*} 	
&\mathbb{E}[F(w_{q+1})]-\mathbb{E}[F(w_{q})]\\
&\leq \mathbb{E}[\langle\nabla F(w_{q}),w_{q+1}-w_{q}\rangle]+\frac{L}{2}\mathbb{E}\|w_{q+1}-w_{q}\|^2\\
&=-\frac{E\eta}{2}\sum_{i\in S_q}\mathbb{E}\|\nabla F^i(w_{q})\|^2+(\frac{L\eta^2E^2}{2}-\frac{E\eta}{2})\\
&*\sum_{i\in S_q}\mathbb{E}\|\sum\limits_{n\in N_{q}^i}\frac{1}{E}\sum_{e=1}^{E}\Tilde{\gamma}_{q,n}^i\nabla F_n^i(w_{q-\tau_q,n,e-1},\xi_{n,e-1})\|^2+\frac{E\eta}{2}(U_3)\\
&\stackrel{a}{\leq} -\frac{E\eta}{2}\sum_{i\in S_q}\mathbb{E}\|\nabla F^i(w_{q})\|^2+\frac{E\eta}{2}(U_3)
\end{align*} 
\end{small}
where $a$ follows because: $\frac{L}{2}\eta^2E^2-\frac{E\eta}{2}<0\Rightarrow \eta<\frac{1}{LE}$

Then, we can obtain:
\begin{small}
\begin{align*} 	
&\mathbb{E}[F(w_{Q+1})]-\mathbb{E}[F(w_{1})]=\sum_{q=1}^Q\mathbb{E}[F(w_{q+1})]-\sum_{q=1}^Q\mathbb{E}[F(w_{q})]\\
&\leq -\frac{E\eta}{2}\sum_{q=1}^Q\mathbb{E}\|\nabla F(w_{q})\|^2+\frac{E\eta}{2}\sum_{q=1}^Q U_3
\end{align*} 
\end{small}
Re-arranging the terms:
\begin{small}
\begin{align*} 	
\frac{E\eta}{2}\sum_{q=1}^Q\mathbb{E}\|\nabla F(w_{q})\|^2\leq \mathbb{E}[F(w_{1})]-\mathbb{E}[F(w_{Q+1})]+\frac{E\eta}{2}\sum_{q=1}^Q U_3
\end{align*} 
\end{small}
Letting $\frac{1}{Q}\sum_{q=1}^Q(\tau_q)^2=\tau$ and
dividing both sides by $\frac{E\eta Q}{2}$
\begin{small}
\begin{align*} 	
&\frac{1}{Q}\sum_{q=1}^Q\mathbb{E}\|\nabla F(w_q)\|^2\\
&\leq \frac{2\mathbb{E}[F(w_{1})]}{E\eta Q}+4L\eta^2E^2\frac{N|K|}{N^*}G(\tau \frac{N|K|}{N^*}+16)\\
&+16L\eta^2E\frac{N}{N^*}|K|\sigma^2+64L\eta^2E^2\frac{N}{N^*}|K|\delta^2
\end{align*} 
\end{small}

\textit{Proof ends.}
\begin{remark}
\textcolor{\attcolor}{Our proof for ${Co\text{-}S}^2{P}$ differs substantially from the traditional FedAvg with full models training. 
\textcircled{\raisebox{-1.0pt}{1}} The model pruning enables all clients to train the subsets of parameters in each round, leading us to introduce new notations such as $N_q^i$ and $N^*$ to represent the client participation in segment-wise training, which directly affects convergence rates. 
\textcircled{\raisebox{-1.0pt}{2}} Unlike FedAvg  aggregating gradients of full models, ${Co\text{-}S}^2{P}$ aggregates only the overlapping gradients corresponding to the pruned submodels. 
\textcircled{\raisebox{-1.0pt}{3}} The masked parameters $\tilde{w}_{0, n}^{l_i}=w_{0, n}^{l_i} \odot \hat{M}_n^{l_i}$ introduces structural differences in the model updates.
\textcircled{\raisebox{-1.0pt}{4}} The semi-asynchronous mechanism of ${Co\text{-}S}^2{P}$
incorporates delay terms, necessitating additional bounding techniques as detailed in Lemma C.2. }   
\end{remark}
Next, by choosing the appropriate convergence rate $\eta$, we can obtain the following corollary. 
\begin{corollary}\label{th:convergence rate}
	Let all assumptions hold. Supposing that the step size $\eta=O(\sqrt{\frac{N^*}{E Q}})$ and $\sigma$ is sufficiently small, when the constant $C>0$ exists, the convergence rate can be expressed as follows:
\begin{small}
\begin{align*} 	
\frac{1}{Q}\sum_{q=1}^Q\mathbb{E}\|\nabla F(w_q)\|^2\leq C(\frac{1}{\sqrt{N^*EQ}}+\frac{1}{Q}).
\end{align*} 
\end{small}
\end{corollary}
\begin{remark}
    Impact of the minimum participation rate $N^*$. The Corollary~\ref{th:convergence rate} shows that the semi-asynchronous aggregation mechanism in ${Co\text{-}S}^2{P}$ can converge to $O(1/\sqrt{N^*EQ})$, while we use the semi-asynchronous aggregate mechanism with the personalized mask trained on local data. The result shows that the larger $N^*$ is, the faster the convergence rate is. 
    Otherwise, it's obvious that $N^*\leq N$. When $N^*=N$, all clients take part in all the communication rounds, which achieves the same convergence rate $O(1/\sqrt{NEQ})$ as asynchronous full client participation \textit{FedAvg}.
\end{remark}

\begin{figure}[t]
  \centering
  \includegraphics[width=0.45\textwidth]{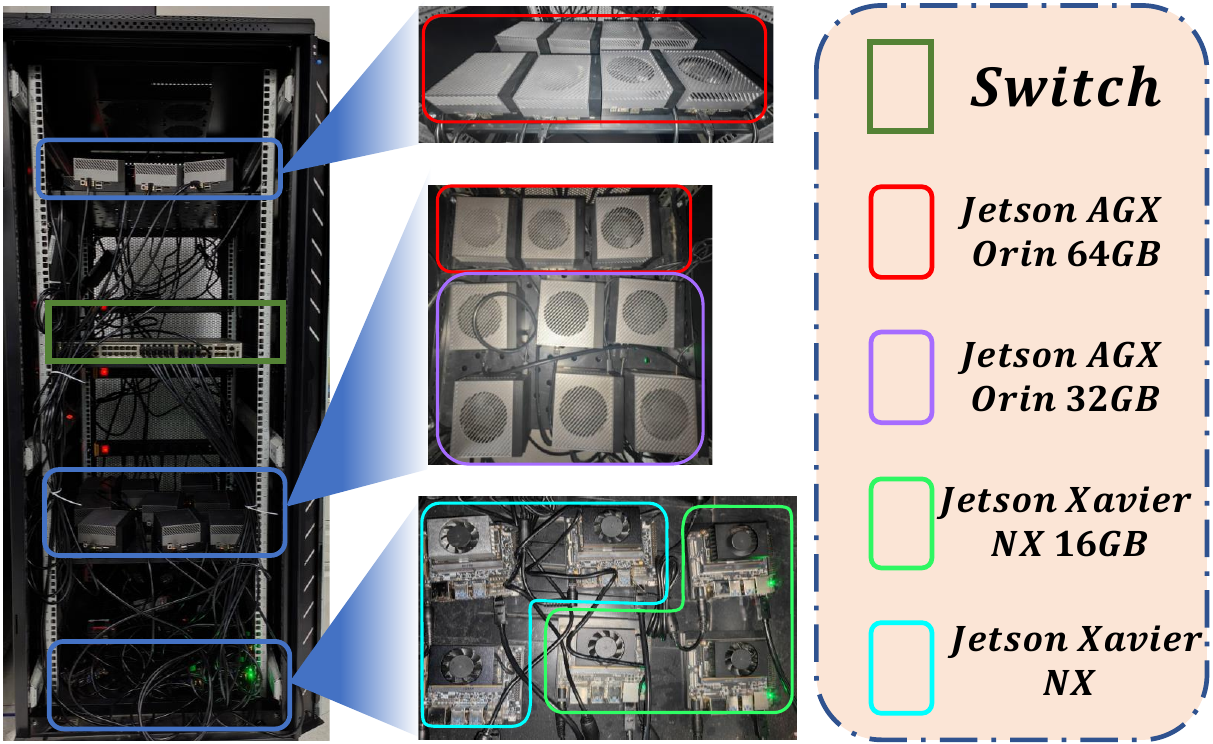}
  \caption{\textcolor{\attcolor}{Overview of the main testbed platform, including 4 types of Jetson devices as the clients and RTX3090 GPU as the server.}}
  \label{fig:system}
\end{figure}

\begin{table}[t]
\caption{Details of three variants of ViT model}
\label{model-variants}
\centering
\scalebox{1.0}{
\begin{tabular}{cccccc}
\toprule
{Model}&{Depth}&{Hidden size}&{MLP size}&{Heads}&{Params} \\
\midrule
ViT-Tiny &8 &512 &2048 &8 &26M \\
ViT-Base &12 &768 &3072 &12 &86M \\
ViT-Base(Ext.) &16 &768 &3072 &12 &0.11B \\
\bottomrule
 \end{tabular}}
\end{table}

\begin{table}[t]
\caption{
\textcolor{\attcolor}{Detailed implementation. 
\textcircled{\raisebox{-1.0pt}{1}}: Jetson Xavier NX \textcircled{\raisebox{-1.0pt}{2}}: Jetson Xavier NX(16GB)
\textcircled{\raisebox{-1.0pt}{3}}: Jetson AGX(32GB)
\textcircled{\raisebox{-1.0pt}{4}}: Jetson AGX(64GB)
}
}
\label{Es}
\centering
\renewcommand{\arraystretch}{0.9}
\scalebox{0.7}{
\begin{tabular}{ccccccccccc}
\toprule
\multirow{2}{*}{Clients}&\multicolumn{3}{c}{E1(ViT-Tiny/ImageNet200)}&\multicolumn{3}{c}{E2(ViT-Base/ImageNet300)}&\multicolumn{3}{c}{E3(ViT-Base(Ext.)/ImageNet300)}\\
\cmidrule(lr){2-4} \cmidrule(lr){5-7} \cmidrule(lr){8-10} 
&{num.}&{$R_n$}&{batch}&{num.}&{$R_n$}&{batch}&{num.}&{$R_n$}&{batch} \\
\midrule
\textcircled{\raisebox{-1.0pt}{1}}&1&0.0625 &8 &1&0.0625 &8 &1&0.0156 &8\\
\cmidrule(lr){2-4} \cmidrule(lr){5-7} \cmidrule(lr){8-10} 
\textcircled{\raisebox{-1.0pt}{2}}&0 &- &- &2 &0.0625 &32 &2 &0.0625 &32\\
\cmidrule(lr){2-4} \cmidrule(lr){5-7} \cmidrule(lr){8-10} 
\textcircled{\raisebox{-1.0pt}{3}}&3 &0.5625 &64 &3 &0.0625 &64 &3 &0.25 &64\\
\cmidrule(lr){2-4} \cmidrule(lr){5-7} \cmidrule(lr){8-10} 
\textcircled{\raisebox{-1.0pt}{4}}&4 &0.5625 &128 &10 &0.5625 &128 &10 &0.5625 &128\\
\bottomrule
 \end{tabular}}
\end{table}

\section{Experiments}\label{expexp}
\subsection{Experimental 
Settings}\label{exp:settings}

\begin{table}[t]
\caption{\textcolor{\attcolor}{Hyperparameter settings for reproduction.
}}
\label{exp:hyperparameter_settings}
  \centering
  \renewcommand{\arraystretch}{1.0}
  \scalebox{0.87}{%
    \begin{tabular}{cccccccccccc}
      \toprule
      \multirow{2}{*}{\textcolor{\attcolor}{Exp ID}}
        & \multirow{2}{*}{\textcolor{\attcolor}{$R$}}
        & \multicolumn{4}{c}{\textcolor{\attcolor}{Mask training}}
        & \multicolumn{4}{c}{\textcolor{\attcolor}{Submodel training}}
        & \multicolumn{2}{c}{\textcolor{\attcolor}{Semi-async.}} \\
      \cmidrule(lr){3-6}\cmidrule(lr){7-10}\cmidrule(lr){11-12}
      & 
        & \textcolor{\attcolor}{$\hat{R}$}
        & \textcolor{\attcolor}{$\hat{E}$}
        & \textcolor{\attcolor}{$\hat{\eta}$}
        & \textcolor{\attcolor}{$\lambda_1$}
        & \textcolor{\attcolor}{$E$}
        & \textcolor{\attcolor}{$\eta$}
        & \textcolor{\attcolor}{$\lambda_2$}
        & \textcolor{\attcolor}{$t$}
        & \textcolor{\attcolor}{$\mu$}
        & \textcolor{\attcolor}{$T_{clk}$} \\
      \midrule
      \textcolor{\attcolor}{E1}
        & \textcolor{\attcolor}{120}
        & \textcolor{\attcolor}{20}
        & \textcolor{\attcolor}{5}
        & \textcolor{\attcolor}{1e-2}
        & \textcolor{\attcolor}{1.0}
        & \textcolor{\attcolor}{5}
        & \textcolor{\attcolor}{5.0e-3}
        & \textcolor{\attcolor}{0.2}
        & \textcolor{\attcolor}{3}
        & \textcolor{\attcolor}{0.5}
        & \textcolor{\attcolor}{60} \\
      \textcolor{\attcolor}{E2}
        & \textcolor{\attcolor}{100}
        & \textcolor{\attcolor}{10}
        & \textcolor{\attcolor}{5}
        & \textcolor{\attcolor}{1e-3}
        & \textcolor{\attcolor}{1.0}
        & \textcolor{\attcolor}{5}
        & \textcolor{\attcolor}{2.5e-4}
        & \textcolor{\attcolor}{0.3}
        & \textcolor{\attcolor}{3}
        & \textcolor{\attcolor}{0.4}
        & \textcolor{\attcolor}{100} \\
      \textcolor{\attcolor}{E3}
        & \textcolor{\attcolor}{100}
        & \textcolor{\attcolor}{10}
        & \textcolor{\attcolor}{5}
        & \textcolor{\attcolor}{1e-2}
        & \textcolor{\attcolor}{1.5}
        & \textcolor{\attcolor}{5}
        & \textcolor{\attcolor}{2.5e-4}
        & \textcolor{\attcolor}{0.3}
        & \textcolor{\attcolor}{3}
        & \textcolor{\attcolor}{0.4}
        & \textcolor{\attcolor}{150} \\
      \bottomrule
    \end{tabular}%
  }
\end{table}

\noindent\textbf{Real-World Testbeds.} To effectively demonstrate the real-world applicability of the proposed collaborative training framework, we utilize a \textit{GeForce RTX 3090 GPU} as the server and 4 types of 16 \textit{Jetson developer kits} as the resource-limited clients to build a main real-world testbed, as shown in Fig.\ref{fig:system}. 


\noindent\textbf{Baselines.}
\textcolor{\attcolor}{To fully evaluate the performance and efficiency of ${Co\text{-}S}^2{P}$, we choose the classical FL algorithm FedAvg~\cite{mcmahan2017communication} and fully asynchronous algorithm FedAsync~\cite{xie2019asynchronous}. 
Additionally, we combine FedAvg with the existing standalone lightweight training works: 
PLATON~\cite{zhang2022platon}, AdaViT~\cite{meng2022adavit} and Q-ViT~\cite{li2022q}, named $FedAvg+$. 
We combine FedAsync with PLATON to evaluate the performance of asynchronous aggregation in the resource-limited scenarios. 
Moreover, we evaluate ${Co\text{-}S}^2{P}$ against several FL algorithms designed for resource-limited scenarios, including FedMeZO~\cite{ling2024convergence}, FedPM~\cite{isik2022sparse}, PruneFL~\cite{jiang2022model}, RAM-Fed~\cite{wang2023theoretical} and FedRolex~\cite{alam2022fedrolex}.}

\noindent\textbf{Backbones and datasets}. We use three ViT~\cite{dosovitskiy2020image} variants with different capabilities, as shown in Tab.\ref{model-variants}. 
We conduct experiments based on \textit{ImageNet1K}~\cite{ILSVRC15} widely used in large-scale model training.
We randomly sample from it 
to get \textit{ImageNet200} with 200 classes and \textit{ImageNet300} with 300 classes respectively. 
For each dataset, we divide it into a testset on the server and a dataset on all clients. We further divide the latter dataset into the dispersed datasets using Dirichlet allocation~\cite{hsu2019measuring} of parameter $\alpha=1.5$. Furthermore, different types of clients have different capacity datasets, as shown in Tab.\ref{Es}. 
For the dispersed datasets in each client, we randomly split into train/test sets with a ratio 8:2.


\noindent\textbf{Evaluation Metrics.} We evaluate the global model on the testset in the server and report the $Top1$ accuracy, ${Top5}$ accuracy, and $F1$-score. 
We also evaluate the personalized masked submodel on the testset in each client and report the average of obtained three values, weighted based on their local dataset sizes.
Furthermore, we introduce a evaluation metric for the real-world testbed, namely \underline{R}esource \underline{U}tilization Rate (RU), calculated as shown below:
\begin{equation}
    RU = \frac{1}{Q}\sum_{q=1}^Q{\frac{\sum_{n=1}^{N_q}{Time(n)}}{N_q \times max(\{Time(1),\cdots,Time(N_q)\})}},
\end{equation}
where 
$Time(n)$ represents the sum of training and communication time of the client n in the $q$-th round.

\noindent\textbf{Implementation.} We deploy ${Co\text{-}S}^2{P}$ and all baselines on the testbed based on the general FL framework NVFlare~\cite{roth2022nvidia}. 
Considering that the unstructured baselines 
fail to work in the resource-limited clients, \textit{we replace the unstructured pruning in the baselines with the structured pruning.} 
\textcolor{\attcolor}{We conduct three groups of real-world experiments, and the detailed configurations are shown in Tab.\ref{Es}. 
Moreover, We use the experimental setting E1 for ablation study and hyperparameter analysis.}

\noindent\textbf{Reproduction details.}
\textcolor{\attcolor}{The details for reproducing three groups of experiments is shown in Tab.\ref{exp:hyperparameter_settings}.
Take E1 as an example, we optimize the local submodels for 120 rounds using early stop strategy and Adan optimizer\cite{xie2022adan}.
And we set the rounds $\hat{R}$, the local epochs $\hat{E}$, the initial learning rate $\hat{\eta}$ and $\lambda_1$ for training width-based structured masks to 20, 5, 1e-2 and 1.0 respectively. 
For the local submodel training in clients, we set the local epochs $E$, the initial learning rate $\eta$, $\lambda_2$ and $t$ to 5, 5.0e-3, 0.2, and 3 respectively.
In the semi-asynchronous aggregation strategy, the minimal ratio $\mu$ and waiting interval $T_{clk}$ are set to 0.5 and 60 seconds, respectively.
Additionally, we set the width pruned rate $R_n^{width}$ equal to depth pruned rate $R_n^{depth}$, both equal $\sqrt{R_n}$, hence the submodels have a balanced learning capacity of both dimensions.}

\begin{table*}[t]
\caption{Performance comparisons on (ViT-Tiny/16)-ImageNet200 and (ViT-Base/16)-ImageNet300. Bold is the optimal result except for FedAvg(FAv) and FedAsync(FAs) with full model training.}
\label{c-performance1_2}
\centering
\renewcommand{\arraystretch}{1.2}
\scalebox{0.99}{
\begin{tabular}{cccccccc|ccccccc}
\toprule
\multirow{3}{*}{Methods}&\multicolumn{7}{c|}{(ViT-Tiny/16)-ImageNet200}&\multicolumn{7}{c}{(ViT-Base/16)-ImageNet300}\\
\cline{2-8} \cline{9-15} 
&\multicolumn{3}{c}{Server(\%)}&\multicolumn{3}{c}{Client(Avg.)(\%)}&\multirow{2}{*}{$RU(\%)$}&\multicolumn{3}{c}{Server(\%)}&\multicolumn{3}{c}{Client(Avg.)(\%)}&\multirow{2}{*}{$RU(\%)$} \\
\cline{2-4} \cline{5-7} \cline{9-11} \cline{12-14}   
&\multicolumn{1}{c}{$Top1$}&\multicolumn{1}{c}{$Top5$}&\multicolumn{1}{c}{$F1$}&\multicolumn{1}{c}{$Top1$}&\multicolumn{1}{c}{$Top5$}&\multicolumn{1}{c}{$F1$}&&\multicolumn{1}{c}{$Top1$}&\multicolumn{1}{c}{$Top5$}&\multicolumn{1}{c}{$F1$}&\multicolumn{1}{c}{$Top1$}&\multicolumn{1}{c}{$Top5$}&\multicolumn{1}{c}{$F1$}\\
\cline{1-15}
FAv(Full)&38.7&63.6&37.3&39.2&64.1&37.1&63.5&68.6&87.2&68.1&64.4&83.2&62.8&60.0\\
\cline{2-15}
\textcolor{\attcolor}{FAs(Full)}
        & \textcolor{\attcolor}{33.5}
        & \textcolor{\attcolor}{59.1}
        & \textcolor{\attcolor}{32.7}
        & \textcolor{\attcolor}{34.0}
        & \textcolor{\attcolor}{60.1}
        & \textcolor{\attcolor}{33.8}
        & \textcolor{\attcolor}{100.0} 
        & \textcolor{\attcolor}{63.7}
        & \textcolor{\attcolor}{82.4}
        & \textcolor{\attcolor}{63.4}
        & \textcolor{\attcolor}{61.8}
        & \textcolor{\attcolor}{81.6}
        & \textcolor{\attcolor}{60.7}
        & \textcolor{\attcolor}{100.0} \\
\cline{1-15}
FAv+PLATON&19.3&42.7&19.1&16.5&37.5&17.4&70.4&37.6&58.2&35.5&34.5&54.2&30.1&74.3\\
\cline{2-9} \cline{9-15}
FAv+AdaViT&17.9&36.1&16.2&14.3&34.5&14.0&67.3&31.9&54.9&32.1&28.6&51.7&29.1&72.7\\
\cline{2-15}
FAv+Q-ViT&17.6&40.1&15.8&14.9&34.8&13.6&70.7&29.7&52.6&29.8&25.6&49.1&26.3&74.5\\
\cline{2-15}
\textcolor{\attcolor}{FAs+PLATON}
        & \textcolor{\attcolor}{14.6}
        & \textcolor{\attcolor}{35.6}
        & \textcolor{\attcolor}{14.2}
        & \textcolor{\attcolor}{13.1}
        & \textcolor{\attcolor}{34.4}
        & \textcolor{\attcolor}{15.7}
        & \textcolor{\attcolor}{100.0}
        & \textcolor{\attcolor}{32.5}
        & \textcolor{\attcolor}{55.7}
        & \textcolor{\attcolor}{31.7}
        & \textcolor{\attcolor}{30.9}
        & \textcolor{\attcolor}{52.2}
        & \textcolor{\attcolor}{30.2}
        & \textcolor{\attcolor}{100.0} \\
\cline{2-15}
\textcolor{\attcolor}{FedMeZO}
        & \textcolor{\attcolor}{16.7}
        & \textcolor{\attcolor}{37.2}
        & \textcolor{\attcolor}{16.3}
        & \textcolor{\attcolor}{15.3}
        & \textcolor{\attcolor}{36.8}
        & \textcolor{\attcolor}{14.9}
        & \textcolor{\attcolor}{67.4} 
        & \textcolor{\attcolor}{24.7}
        & \textcolor{\attcolor}{45.1}
        & \textcolor{\attcolor}{23.8}
        & \textcolor{\attcolor}{24.2}
        & \textcolor{\attcolor}{44.8}
        & \textcolor{\attcolor}{23.6}
        & \textcolor{\attcolor}{73.1} \\
\cline{2-15}
FedPM&16.2&38.7&15.5&14.2&34.0&12.9&71.8&26.4&49.3&27.2&23.7&44.8&24.6&76.2\\
\cline{2-15}
PruneFL&21.8&42.2&19.1&20.2&39.1&17.1&69.2&42.1&65.8&41.3&37.1&61.6&36.2&79.6\\
\cline{2-15}
RAM-Fed&23.6&46.0&21.5&19.4&41.2&19.3&69.9&43.4&66.7&42.0&36.8&59.6&35.3&80.5\\
\cline{2-15}
FedRolex&24.3&48.0&23.1&20.7&45.8&20.5&72.5&44.2&67.5&43.4&38.4&62.1&38.6&82.4\\
\cline{1-15}
$\text{Co-S}^2\text{P}$&\textbf{33.1}&\textbf{59.2}&\textbf{31.9}&\textbf{31.8}&\textbf{56.6}&\textbf{30.5}&{87.9}&\textbf{50.6}&\textbf{76.4}&\textbf{48.9}&\textbf{47.2}&\textbf{72.5}&\textbf{45.5}&{92.4}\\
\cline{2-15}
$\text{Co-S}^2\text{P}$(Async.)&27.7&53.2&27.3&26.7&49.8&26.7&\textbf{100.0}&47.6&71.9&46.3&44.1&68.4&42.6&\textbf{100.0}\\
\bottomrule
 \end{tabular}}
\end{table*}

\vspace{-0.5em}
\subsection{Overall Performance}
\begin{figure*}[t]
    \centering
    \vspace{-0.7em}
    \begin{minipage}[t]{1.0\linewidth}
    \centering
        \begin{tabular}{@{\extracolsep{\fill}}c@{}c@{}c@{}@{\extracolsep{\fill}}}
            \includegraphics[width=0.32\linewidth]{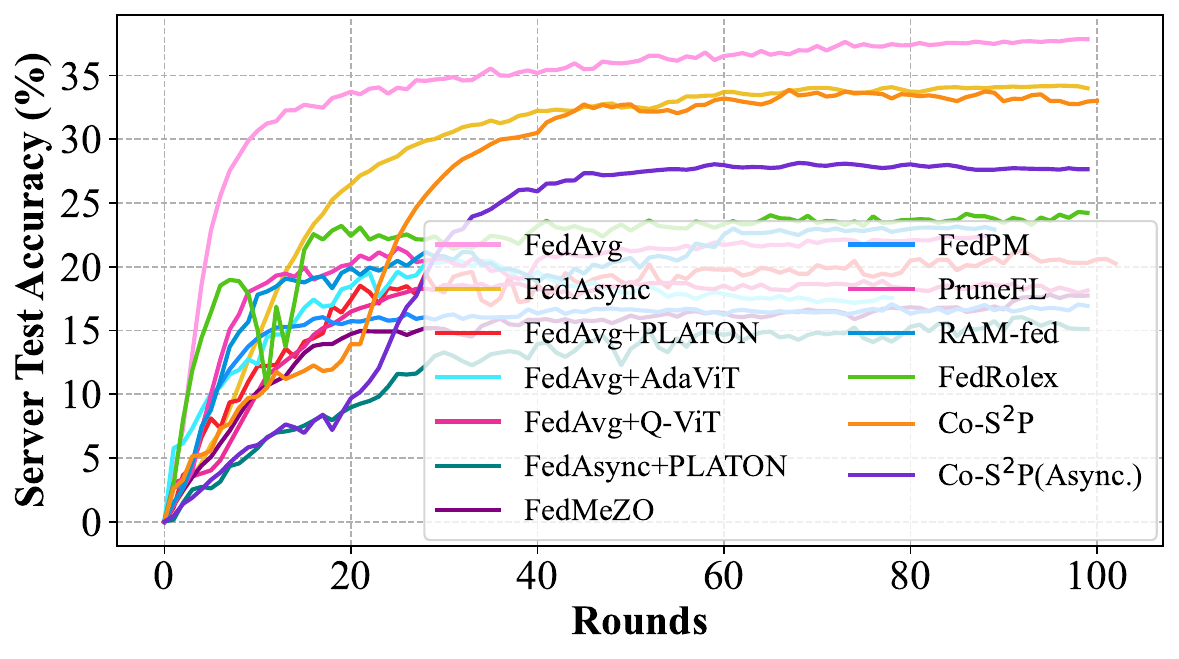} &
            \includegraphics[width=0.32\linewidth]{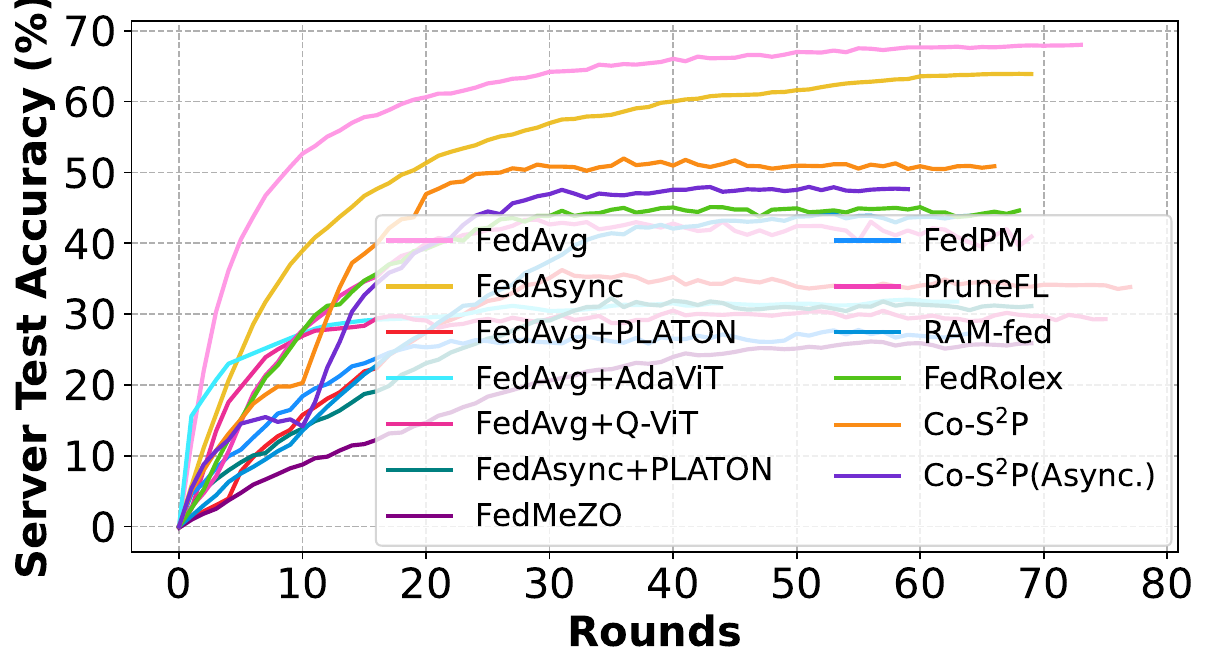}&
            \includegraphics[width=0.32\linewidth]{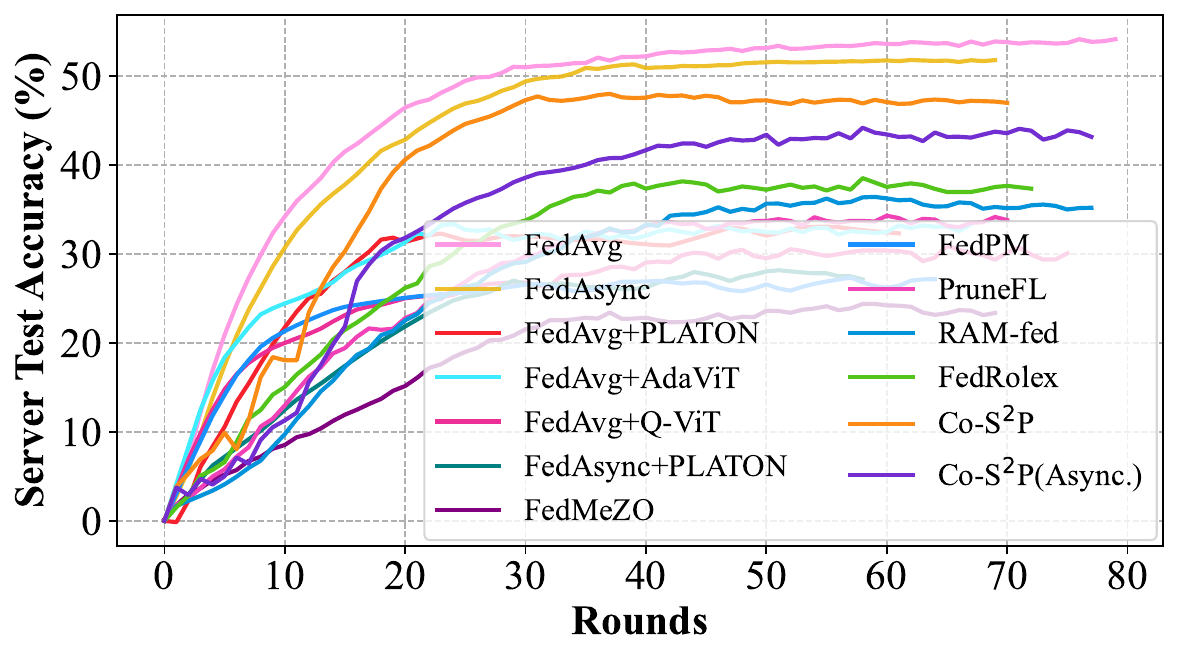}\\
            (a) (ViT-Tiny)/(ImageNet200) & (b) (ViT-Base)/(ImageNet300) & (c) (ViT-Base(ext.))/(ImageNet300)\\
        \end{tabular}
    \end{minipage}
    \caption{Convergence curves of FedAvg, all baselines and ours algorithm in three groups of experiment. ${Co\text{-}S}^2{P}$ achieves better performance and comparable or even better convergence speeds than baselines.}
    \label{fig:conver_curve}
 \end{figure*}

\textbf{Server's Performance.} 
As shown in Fig.\ref{fig:conver_curve}, Tab.\ref{c-performance1_2} and Tab.\ref{c-performance3}, ${Co\text{-}S}^2{P}$ outperforms all baselines in terms of global performance across three groups of experiments: 
(1) For \textit{Top1} accuracy, ${Co\text{-}S}^2{P}$ achieves 8.8\%/6.4\%/8.7\% higher global model performance compared to the closest baseline FedRolex on three groups of experiments respectively. For \textit{Top5} accuracy, ${Co\text{-}S}^2{P}$ improves 11.2\%/8.9\%/9.3\%, respectively. For \textit{F1-score}, ${Co\text{-}S}^2{P}$ improves 8.8\%/5.5\%/8.7\% respectively. These demonstrates that ${Co\text{-}S}^2{P}$ achieves better generalization in resource-limited scenarios. 
(2) For \textit{Top1} accuracy, the $FedAvg+$ baselines significantly decrease $2.0\% \sim 14.5\%$ compared to the federated resource-adaptive methods, indicating that the mixed heterogeneity of collaborative training scenarios has severe implications for model training. 
\textcolor{\attcolor}{(3) Compared with ${Co\text{-}S}^2{P}$, the \textit{Top1} accuracy of FedMeZO is reduced by 16.4\%/25.9\%/23.8\% respectively, 
demonstrating that FedMeZO is not suitable for full-parameter training of large models.
(4) Considering \textit{Top1} accuracy, ${Co\text{-}S}^2{P}$ achieves 18.5\%/18.1\%/20.6\% higher improvement compared to FedAsync+PLATON on three groups of experiments respectively. This highlights that ${Co\text{-}S}^2{P}$ effectively balances resource utilization and global model's performance.}
 
\textbf{Clients' Average Performance.} As shown in Tab.\ref{c-performance1_2} and Tab.\ref{c-performance3}, ${Co\text{-}S}^2{P}$ outperforms all baselines in terms of weighted average performance of the clients in three groups of experiments: 
(1) For \textit{Top1} accuracy, ${Co\text{-}S}^2{P}$ achieve 11.1\%/8.8\%/10.1\% higher personalized submodel performance compared to the closest approach on three groups of experiments respectively. For Top5 accuracy, ${Co\text{-}S}^2{P}$ improves 10.6\%/10.4\%/9.0\% respectively. For \textit{F1-score}, ${Co\text{-}S}^2{P}$ improves 10.5\%/6.9\%/9.8\% respectively. These demonstrate that our framework obtain a better personalized mask for each client in resource-limited scenarios.
(2) Comparing the difference between global and local performance, ${Co\text{-}S}^2{P}$ has closer gap than all baselines. This is because the trainable masks better adapt to local data distributions and the depth-pruned submodels in resource-limited clients.
\textcolor{\attcolor}{(3) Considering \textit{Top1} accuracy, ${Co\text{-}S}^2{P}$ achieves 18.7\%/16.3\%/20.1\% higher the average convergence performance of submodels compared to FedAsync+PLATON on three groups of experiments respectively, which demonstrates that the semi-asynchronous collaborative training framework ${Co\text{-}S}^2{P}$ effectively enables submodels to achieve a superior balance between resource efficiency and personalized performance.
(4) $Co\text{-}S^2P$ achieves  $11.6 \%/10.1 \%$/$14.7 \%$ higher the average convergence performance of submodels compared to PruneFL across three different experimental groups respectively. These results highlight that the trainable masks can better adapt to local distributions and improve performance.}

\begin{figure*}[thb!]
    \centering
    \begin{minipage}[t]{1.0\linewidth}
    \vspace{-0.7em}
    \centering
        \begin{tabular}{@{\extracolsep{\fill}}c@{}c@{}c@{}@{\extracolsep{\fill}}}
            \includegraphics[width=0.32\linewidth]{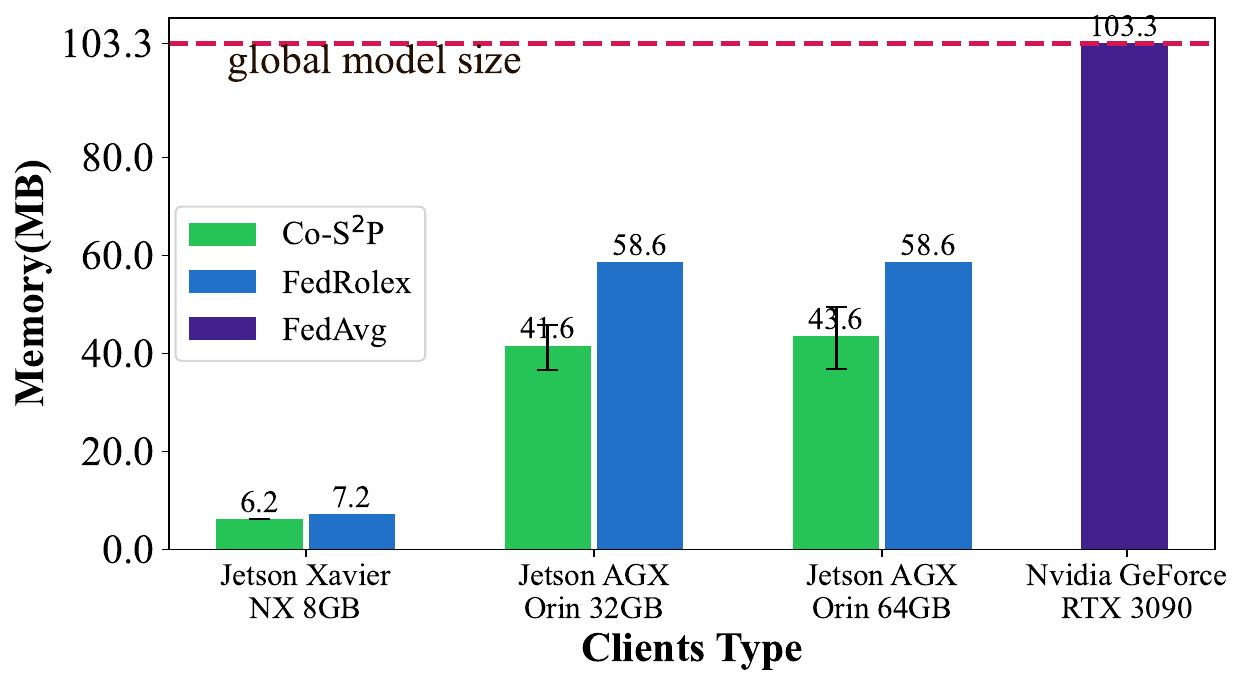} &
            \includegraphics[width=0.32\linewidth]{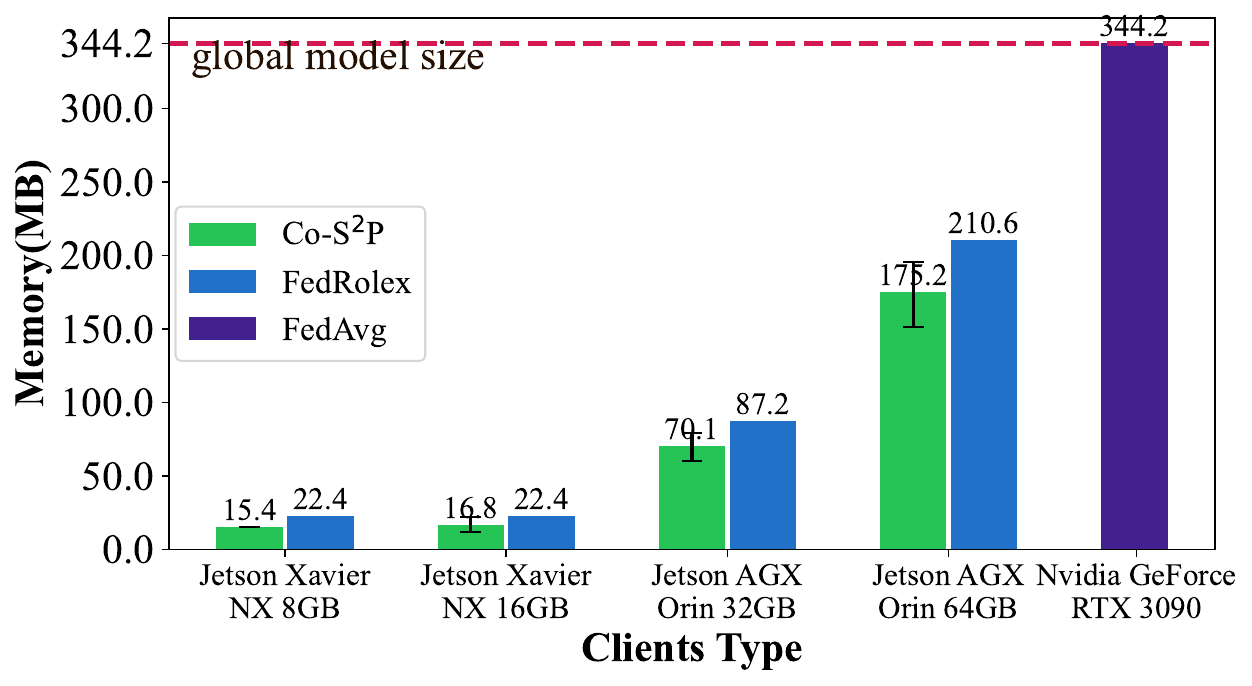}&
            \includegraphics[width=0.32\linewidth]{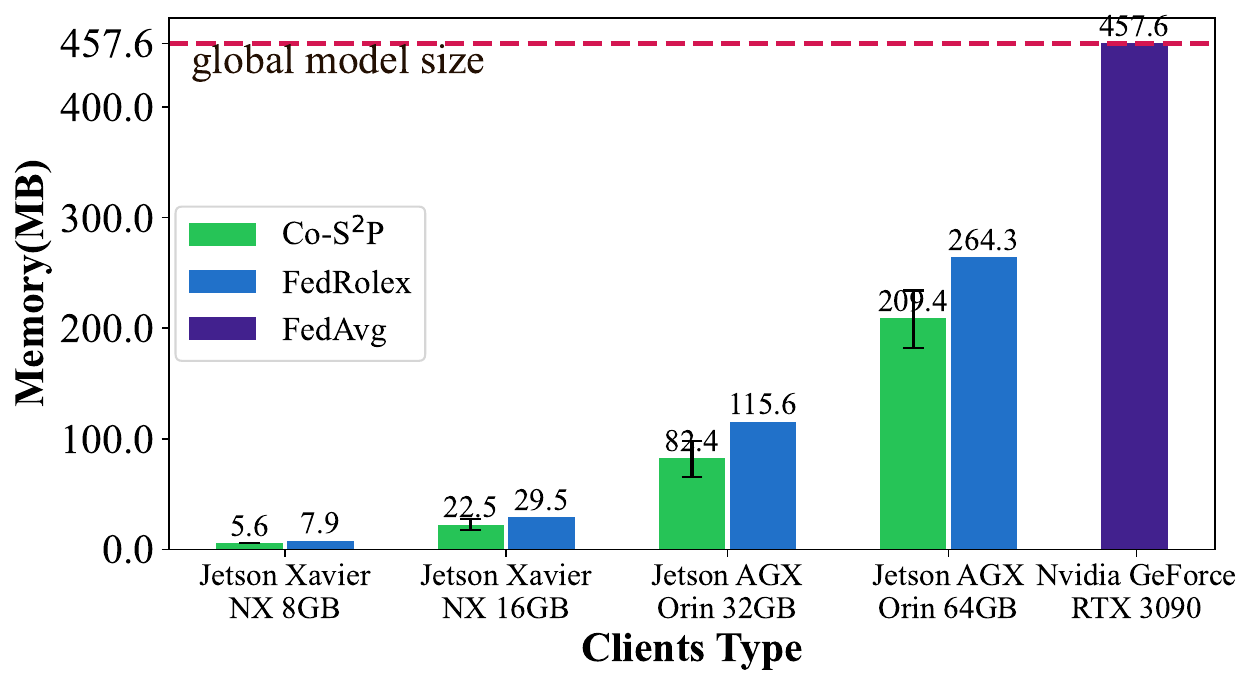}\\
            (a) (ViT-Tiny)/(ImageNet200) & (b) (ViT-Base)/(ImageNet300) & (c) (ViT-Base(ext.))/(ImageNet300)\\
        \end{tabular}
    \end{minipage}
    \caption{Comparison of model memory usage. We use FedRolex with the closest performance for comparison. The red horizontal line represents the memory usage during global model training. ${Co\text{-}S}^2{P}$ reduces memory consumption by about \textbf{22\%} across all resource-limited devices.}
    \label{fig:mem}
 \end{figure*}

\begin{figure*}[thb!]
  \vspace{-0.7em}
    \centering
    \begin{minipage}[t]{1.0\linewidth}
    \centering
        \begin{tabular}{@{\extracolsep{\fill}}c@{}c@{}c@{}@{\extracolsep{\fill}}}
            \includegraphics[width=0.32\linewidth]{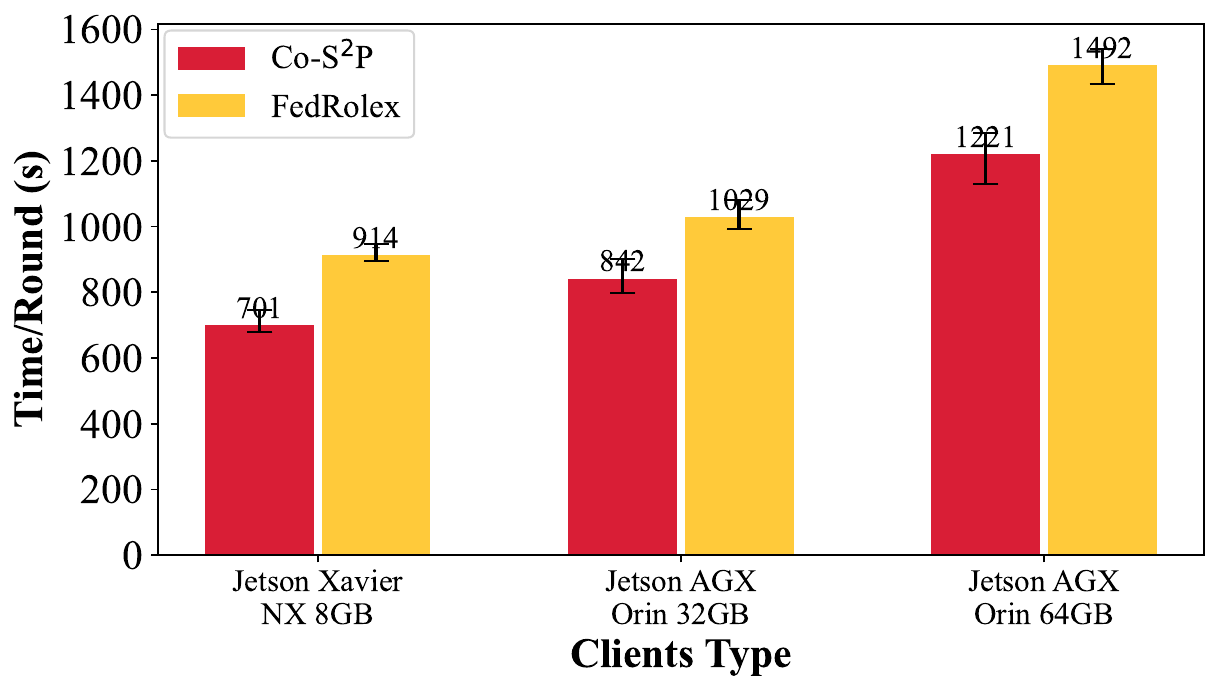} &
            \includegraphics[width=0.32\linewidth]{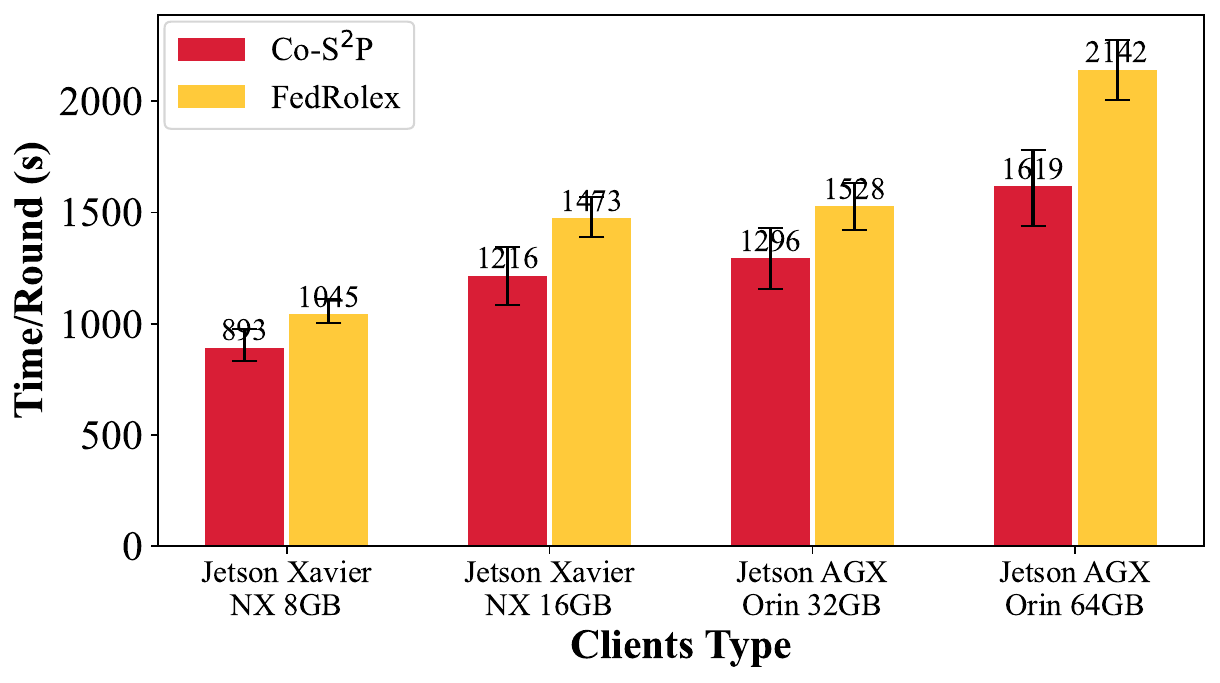}&
            \includegraphics[width=0.32\linewidth]{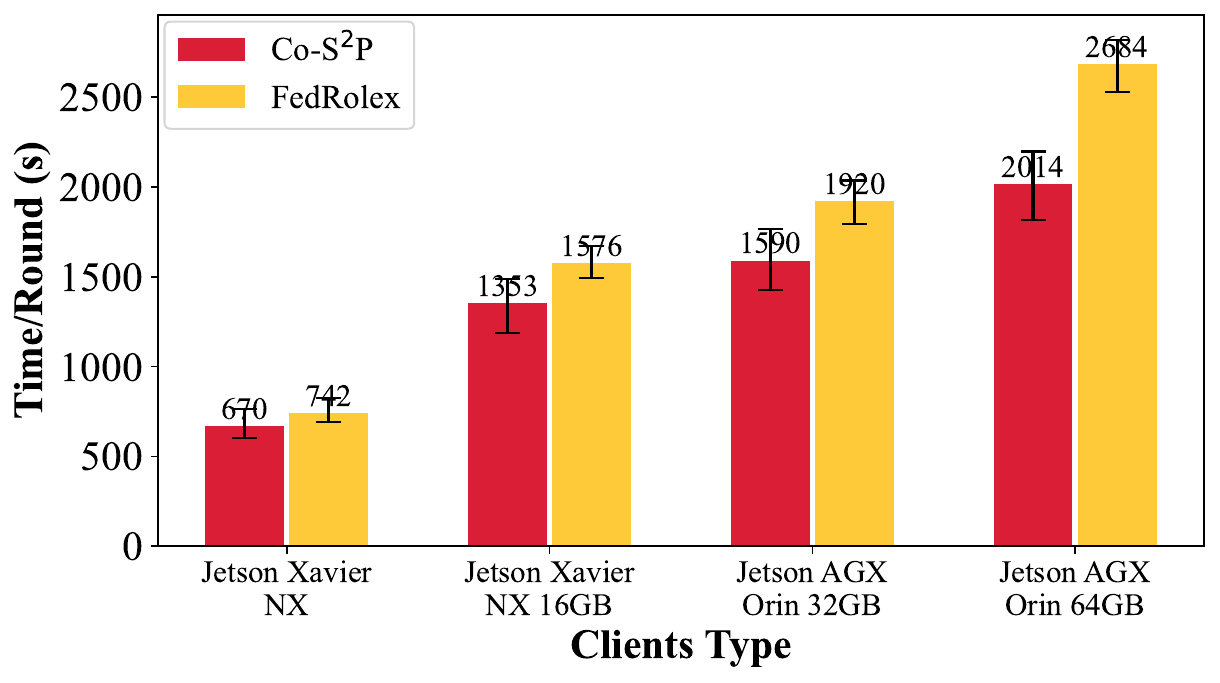}\\
            (a) (ViT-Tiny)/(ImageNet200) & (b) (ViT-Base)/(ImageNet300) & (c) (ViT-Base(ext.))/(ImageNet300)\\
        \end{tabular}
    \end{minipage}
    \caption{Comparison of training time per round. ${Co\text{-}S}^2{P}$ reduces training time per round by about \textbf{24\%} across all resource-limited devices.}
    \label{fig:time}
 \end{figure*}

\textbf{Efficiency Comparison.} As shown in Fig.\ref{fig:mem} and Fig.\ref{fig:time}, ${Co\text{-}S}^2{P}$ outperforms all baselines in terms of efficiency in three groups of experiments: 
(1) Compared to FedRolex, ${Co\text{-}S}^2{P}$ reduces memory consumption by about \textbf{22\%} and training time per round by about \textbf{24\%} on all resource-limited devices, demonstrating the superiority of the structured pruning in memory and computation efficiency. (2) For resource utilization rate, ${Co\text{-}S}^2{P}$ achieves \textbf{1.4$\times$}, \textbf{1.5$\times$}, and \textbf{1.5$\times$} improvements compared to FedAvg and 1.2$\times$, 1.1$\times$, and 1.2$\times$ improvements compared to FedRolex across the three groups of experiments. This demonstrates that the proposed semi-asynchronous aggregation strategy can effectively improve the resource utilization during large-scale model training.


\begin{table*}[t]
\caption{Performance comparison on (ViT-Base(Ext.)/16)-ImageNet300 with parameters up to 0.11 billion. 
}
\label{c-performance3}
\centering
\renewcommand{\arraystretch}{0.1}
\scalebox{1.0}{
\begin{tabular}{cccccccc}
\toprule
\multirow{2}{*}{Methods}&\multicolumn{3}{c}{Server}&\multicolumn{3}{c}{Client(Avg.)}&\multirow{2}{*}{$RU(\%)$} \\
\cmidrule(lr){2-4} \cmidrule(lr){5-7}  &$Top1(\%)$&$Top5(\%)$&$F1\text{-}score(\%)$&$Top1(\%)$&$Top5(\%)$&$F1\text{-}score(\%)$&\\
\midrule
FedAvg(Full)&56.4&78.3&55.8&50.4&73.0&49.9&61.4\\ 
\cmidrule(lr){2-8}
\textcolor{\attcolor}{FedAsync(Full)}
        & \textcolor{\attcolor}{51.7}
        & \textcolor{\attcolor}{74.9}
        & \textcolor{\attcolor}{50.6}
        & \textcolor{\attcolor}{47.8}
        & \textcolor{\attcolor}{69.7}
        & \textcolor{\attcolor}{46.9}
        & \textcolor{\attcolor}{100.0} \\
\cmidrule(lr){1-8}
FedAvg+PLATON&32.3&55.3&31.6&30.7&51.8&31.6&75.1\\
\cmidrule(lr){2-8}
FedAvg+AdaViT&31.4&54.3&31.9&30.8&51.4&29.3&74.4\\ 
\cmidrule(lr){2-8}
FedAvg+Q-ViT&30.7&53.2&30.2&28.1&52.1&27.6&74.1\\ 
\cmidrule(lr){2-8}
      \textcolor{\attcolor}{FedAsync+PLATON}
        & \textcolor{\attcolor}{26.3}
        & \textcolor{\attcolor}{48.5}
        & \textcolor{\attcolor}{25.8}
        & \textcolor{\attcolor}{25.6}
        & \textcolor{\attcolor}{47.6}
        & \textcolor{\attcolor}{25.1}
        & \textcolor{\attcolor}{100.0} \\
      \cmidrule(lr){2-8}
      \textcolor{\attcolor}{FedMeZO}
        & \textcolor{\attcolor}{23.1}
        & \textcolor{\attcolor}{43.0}
        & \textcolor{\attcolor}{23.4}
        & \textcolor{\attcolor}{22.7}
        & \textcolor{\attcolor}{41.8}
        & \textcolor{\attcolor}{22.5}
        & \textcolor{\attcolor}{72.5} \\
\cmidrule(lr){2-8}
FedPM&27.3&52.2&26.3&25.4&49.1&24.1&73.6\\ 
\cmidrule(lr){2-8}
PruneFL&34.3&58.4&32.8&31.0&54.4&30.4&76.9\\ 
\cmidrule(lr){2-8}
RAM-Fed&36.4&60.8&35.5&30.4&53.1&30.1&77.5\\  
\cmidrule(lr){2-8}
FedRolex&38.1&63.2&37.3&35.6&58.5&34.7&78.2\\ 
\cmidrule(lr){1-8}
$\text{Co-S}^2\text{P}$&\textbf{46.9}&\textbf{72.5}&\textbf{46.0}&\textbf{45.7}&\textbf{67.5}&\textbf{44.5}&90.6\\
\cmidrule(lr){2-8}
$\text{Co-S}^2\text{P}$(Async.)&43.4&68.2&43.6&41.3&65.6&41.5&\textbf{100.0}\\
\bottomrule
 \end{tabular}}
\end{table*}

 
\subsection{Ablation Study}
\label{exp:ablation}
\begin{table}[t]
\caption{\textcolor{\attcolor}{Effectiveness of 
trainable mask policies, self-distillation mechanism and semi-asynchronous weight aggregation.}}
\label{table:ablation}
\centering
\renewcommand{\arraystretch}{0.9}
\scalebox{0.87}{
\begin{tabular}{ccccccccc}
\toprule
      \multicolumn{2}{c}{\textcolor{\attcolor}{Trainable Mask}}
        & \multirow{2}{*}[-0.5ex]{\textcolor{\attcolor}{\makecell[c]{Self-\\dist.}}}
        & \multirow{2}{*}[-0.5ex]{\textcolor{\attcolor}{\makecell[c]{Semi-\\async.}}}
        & \multicolumn{2}{c}{\textcolor{\attcolor}{Server (\%)}}
        & \multicolumn{2}{c}{\textcolor{\attcolor}{Client (Avg.) (\%)}}
        & \multirow{2}{*}{\textcolor{\attcolor}{$RU(\%)$}} \\
      \cmidrule(lr){1-2}\cmidrule(lr){5-6}\cmidrule(lr){7-8}
      \textcolor{\attcolor}{Linear}
        & \textcolor{\attcolor}{MSA}
        & & 
        & \textcolor{\attcolor}{$Top1$}
        & \textcolor{\attcolor}{$Top5$}
        & \textcolor{\attcolor}{$Top1$}
        & \textcolor{\attcolor}{$Top5$}
        & \\
      \midrule
      \textcolor{\attcolor}{\ding{55}}
        & \textcolor{\attcolor}{\ding{55}}
        & \textcolor{\attcolor}{\ding{55}}
        & \textcolor{\attcolor}{\checkmark}
        & \textcolor{\attcolor}{18.4}
        & \textcolor{\attcolor}{41.2}
        & \textcolor{\attcolor}{17.4}
        & \textcolor{\attcolor}{40.9}
        & \textcolor{\attcolor}{88.1} \\
      \cmidrule(lr){1-4}\cmidrule(lr){5-9}
      \textcolor{black}{\ding{55}}
        & \textcolor{black}{\ding{55}}
        & \textcolor{black}{\checkmark}
        & \textcolor{black}{\checkmark}
        & \textcolor{black}{22.5}
        & \textcolor{black}{45.3}
        & \textcolor{black}{18.9}
        & \textcolor{black}{40.2}
        & \textcolor{\attcolor}{87.8} \\
      \cmidrule(lr){1-4}\cmidrule(lr){5-9}
      \textcolor{black}{\ding{55}}
        & \textcolor{black}{\checkmark}
        & \textcolor{black}{\checkmark}
        & \textcolor{black}{\checkmark}
        & \textcolor{black}{26.5}
        & \textcolor{black}{51.6}
        & \textcolor{black}{24.3}
        & \textcolor{black}{47.2}
        & \textcolor{\attcolor}{88.0} \\
      \cmidrule(lr){1-4}\cmidrule(lr){5-9}
      \textcolor{black}{\checkmark}
        & \textcolor{black}{\ding{55}}
        & \textcolor{black}{\checkmark}
        & \textcolor{black}{\checkmark}
        & \textcolor{black}{27.8}
        & \textcolor{black}{55.4}
        & \textcolor{black}{25.3}
        & \textcolor{black}{50.9}
        & \textcolor{\attcolor}{87.9} \\
      \cmidrule(lr){1-4}\cmidrule(lr){5-9}
      \textcolor{black}{\checkmark}
        & \textcolor{black}{\checkmark}
        & \textcolor{black}{\ding{55}}
        & \textcolor{black}{\checkmark}
        & \textcolor{black}{28.2}
        & \textcolor{black}{56.1}
        & \textcolor{black}{22.3}
        & \textcolor{black}{47.5}
        & \textcolor{\attcolor}{87.6} \\
      \cmidrule(lr){1-4}\cmidrule(lr){5-9}
      \textcolor{\attcolor}{\checkmark}
        & \textcolor{\attcolor}{\checkmark}
        & \textcolor{\attcolor}{\checkmark}
        & \textcolor{\attcolor}{\ding{55}}
        & \textcolor{\attcolor}{35.2}
        & \textcolor{\attcolor}{62.1}
        & \textcolor{\attcolor}{34.0}
        & \textcolor{\attcolor}{61.1}
        & \textcolor{\attcolor}{78.1} \\
      \cmidrule(lr){1-4}\cmidrule(lr){5-9}
      \textcolor{\attcolor}{\checkmark}
        & \textcolor{\attcolor}{\checkmark}
        & \textcolor{\attcolor}{\ding{55}}
        & \textcolor{\attcolor}{\ding{55}}
        & \textcolor{\attcolor}{31.5}
        & \textcolor{\attcolor}{57.8}
        & \textcolor{\attcolor}{27.4}
        & \textcolor{\attcolor}{48.9}
        & \textcolor{\attcolor}{78.4} \\
      \cmidrule(lr){1-4}\cmidrule(lr){5-9}
      \textcolor{\attcolor}{\checkmark}
        & \textcolor{\attcolor}{\checkmark}
        & \textcolor{\attcolor}{\checkmark}
        & \textcolor{\attcolor}{\checkmark}
        & \textcolor{\attcolor}{33.1}
        & \textcolor{\attcolor}{59.2}
        & \textcolor{\attcolor}{31.8}
        & \textcolor{\attcolor}{56.6}
        & \textcolor{\attcolor}{87.9} \\
      \bottomrule
 \end{tabular}}
 \vspace{-0.3em}
\end{table}

\textbf{Effectiveness of trainable mask policies.} We validate that the trainable mask strategy is effective in improving global performance and generating personalized masks. We replace the trainable masking strategies with randomly mask strategies and report the results in Tab.\ref{table:ablation}. We observe that replacing any set of trainable masks with randomized ones decreases the convergence performances. 
With MSA/Linear/Both Random, $Co\text{-}S^2P$ improves global model accuracy by 5.3\%/6.6\%/10.6\%, and the average accuracy of submodels by 6.5\%/7.5\%/12.9\% in \textit{Top1} accuracy. 
These confirm the effectiveness of the trainable mask strategy.


\textbf{Effectiveness of self-distillation mechanism.} We validate that the self-distillation mechanism helps shallow submodels on resource-limited clients to learn high-level knowledge. 
As shown in Tab.\ref{table:ablation}, $Co\text{-}S^2P$ improves Top1 accuracy by 4.9\% for the global performance and by 9.5\% for the average performance of submodels. 
This demonstrates the effectiveness of self-distillation in enabling shallow models on resource-limited clients to obtain more complex and high-level knowledge.
\textcolor{\attcolor}{Moreover, the overall resource utilization remains nearly unchanged with or without the self-distillation mechanism, further confirming that it does not impose additional overhead.}

\textcolor{\attcolor}{\textbf{Effectiveness of modules' combination.} 
We further conduct the ablation combinations of the three proposed modules to demonstrate that they are effectively combined to improve convergence performance while enhancing resource utilization.
As shown in Tab.\ref{table:ablation}, the trainable mask algorithm and the self-distillation mechanism individually contribute to improvements of 9.8\% and 4.1\% in \textit{Top1} accuracy, respectively. 
Notably, the overall framework achieves a 14.7\% increase—exceeding the sum of the individual gains (9.8\% + 4.1\%).
These illustrates that the benefits brought by the trainable masks and self-distillation are synergistic rather than merely additive.
Additionally, the improvements from three modules exhibit complementary effects: the trainable mask and self-distillation primarily enhance accuracy, while the semi-asynchronous strategy significantly boosts resource utilization.}

\subsection{Hyperparameter Analysis}

\begin{figure}[t]
  \centering
  \includegraphics[width=0.5\textwidth]{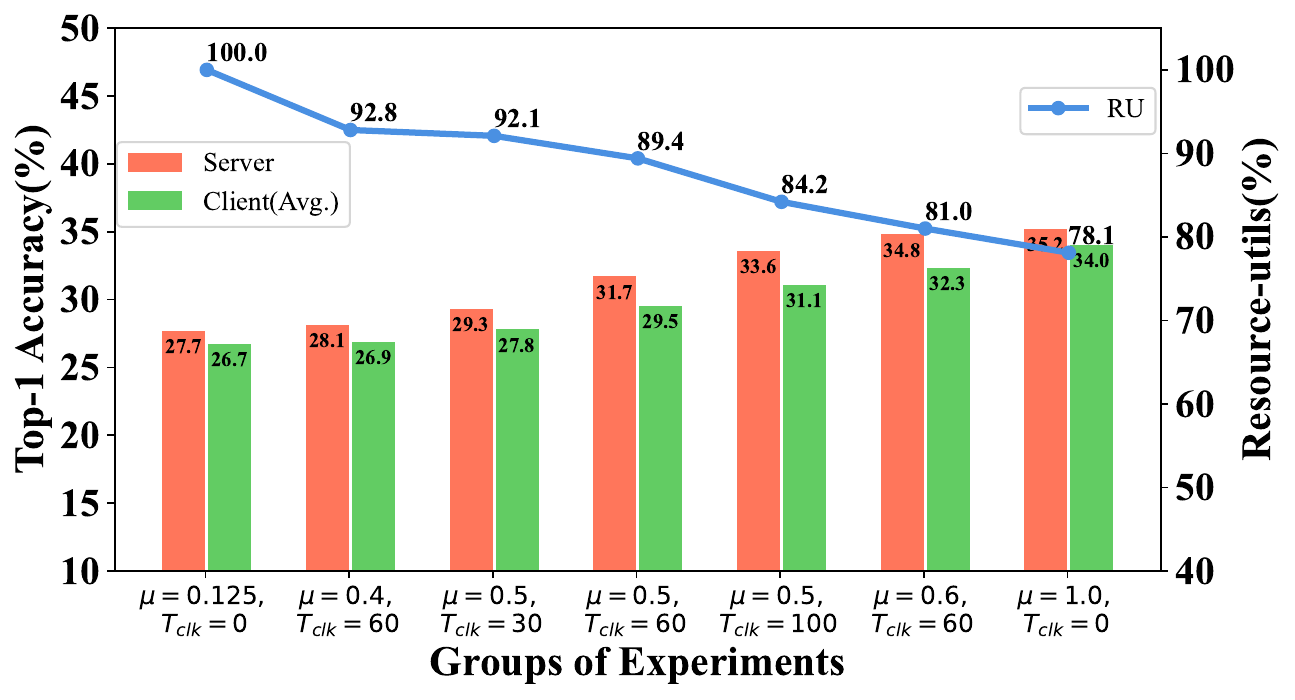}
  \vspace{-2em}
  \caption{Semi-asynchronous aggregation strategy with different $\mu$ and $T_{clk}(s)$.}
  \label{semi-async_exp}
\end{figure}

\begin{figure}[t]
    \centering
    \begin{minipage}[t]{1.0\linewidth}
    \centering
        \begin{tabular}{@{\extracolsep{\fill}}c@{}c@{}c@{}@{\extracolsep{\fill}}}
            \includegraphics[width=0.5\linewidth]{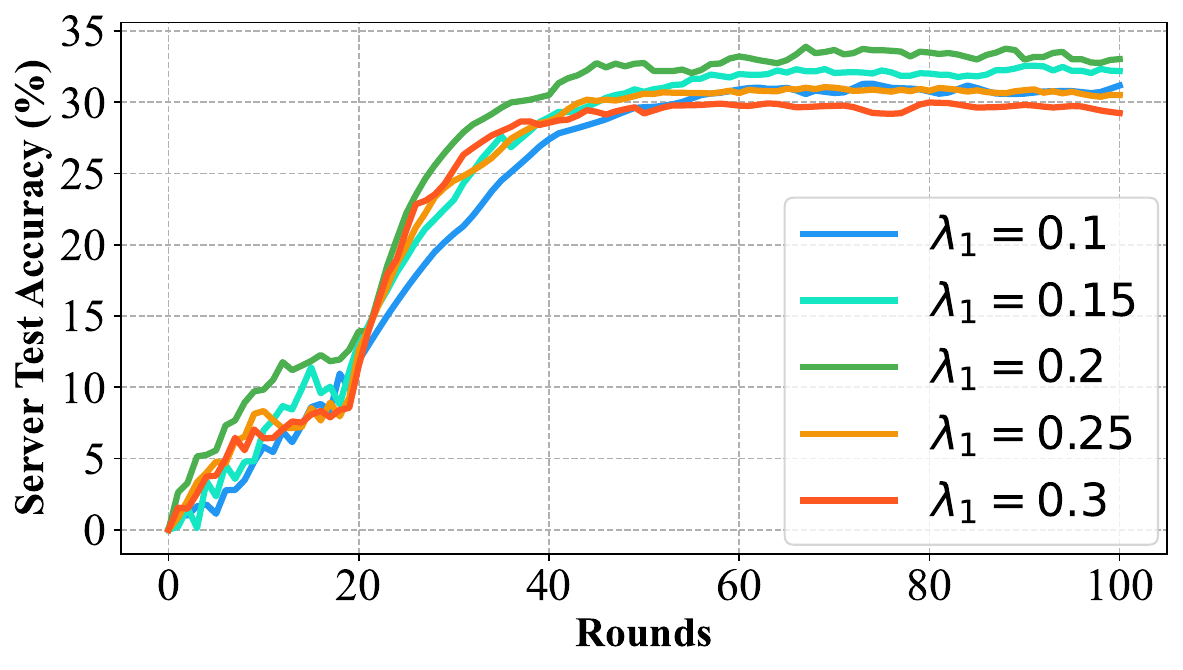} &
            \includegraphics[width=0.5\linewidth]{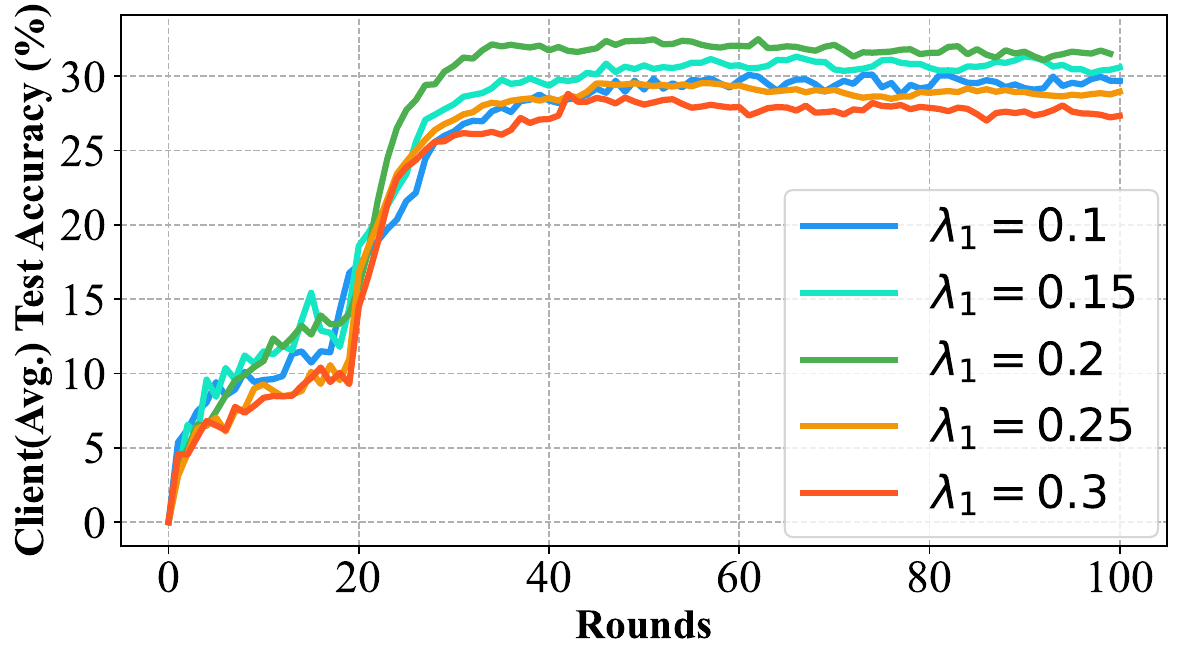}&\\
            \textcolor{\attcolor}{(a) Test in Server} & \textcolor{\attcolor}{(b) Test in Clients(Avg.)}\\
        \end{tabular}
    \end{minipage}
    \caption{\textcolor{\attcolor}{Impact of $\lambda_1$ for training the width structured masks.}}
    \label{fig:lambda1}
 \end{figure}

\begin{figure}[t]
    \centering
    \begin{minipage}[t]{1.0\linewidth}
    \centering
        \begin{tabular}{@{\extracolsep{\fill}}c@{}c@{}c@{}@{\extracolsep{\fill}}}
            \includegraphics[width=0.5\linewidth]{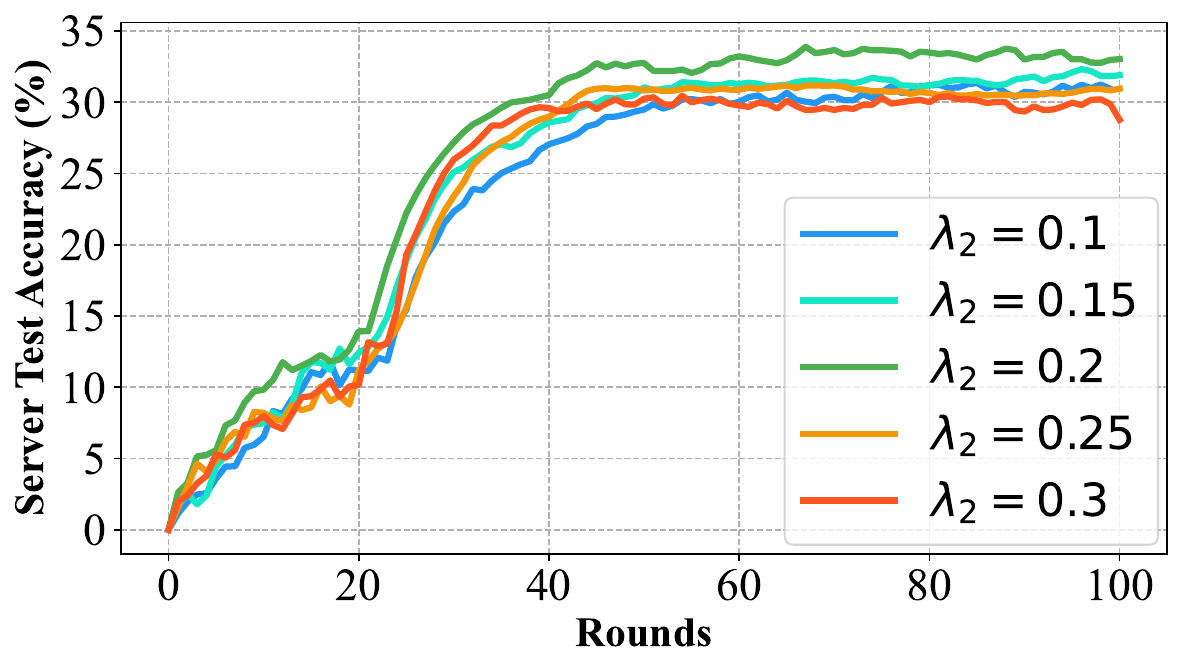} &
            \includegraphics[width=0.5\linewidth]{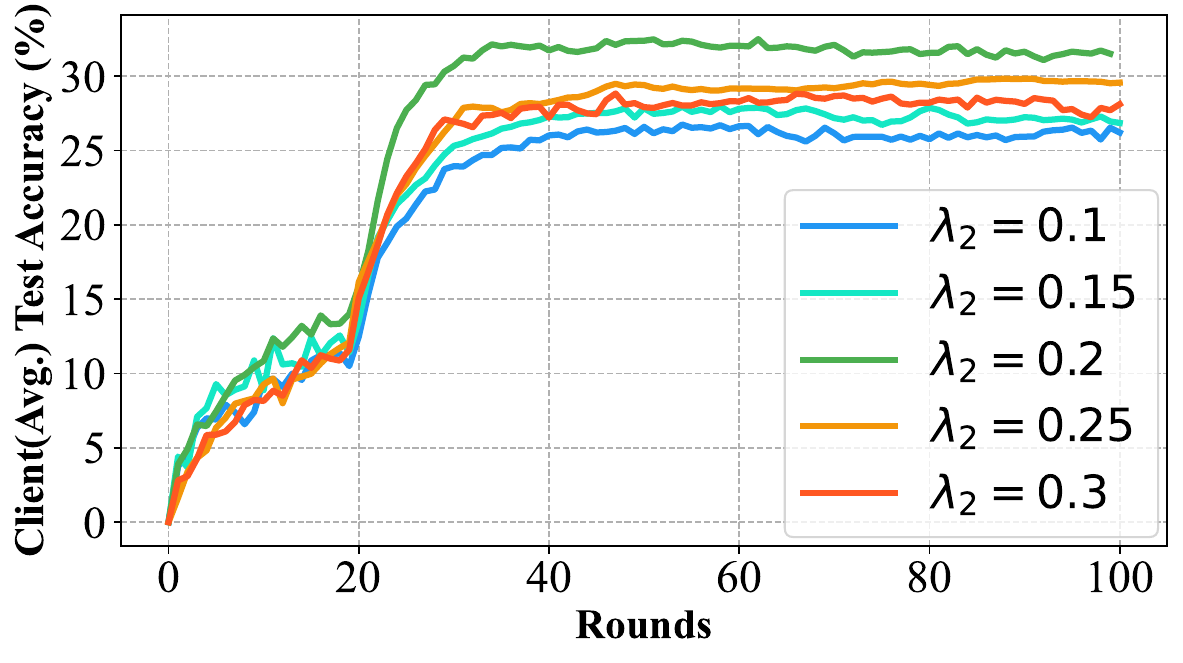}&\\
            \textcolor{\attcolor}{(a) Test in Server} & \textcolor{\attcolor}{(b) Test in Clients(Avg.)}\\
        \end{tabular}
    \end{minipage}
    \caption{\textcolor{\attcolor}{Impact of $\lambda_2$ within the structured-pruned submodel training stage.}}
    \label{fig:lambda2}
 \end{figure}

\begin{table}[t!]
\caption{Impact of different ratio of $R_n^{width}$ and $R_n^{depth}$.}
\label{table:pruneratio}
\centering
\renewcommand{\arraystretch}{0.7}
\scalebox{1.0}{
\begin{tabular}{ccccccc}
\toprule
      \multirow{2}{*}{\makecell[c]{${R_n^{width}}/{R_n^{depth}}$}}
        & \multicolumn{3}{c}{Server(\%)}
        & \multicolumn{3}{c}{Client(Avg.)(\%)} \\
      \cmidrule(lr){2-4}\cmidrule(lr){5-7}
      & $Top1$ & $Top5$ & $F1$
      & $Top1$ & $Top5$ & $F1$ \\
      \midrule
      \textcolor{\attcolor}{0 (only depth)}
        & \textcolor{\attcolor}{23.8}
        & \textcolor{\attcolor}{46.5}
        & \textcolor{\attcolor}{23.2}
        & \textcolor{\attcolor}{22.7}
        & \textcolor{\attcolor}{44.1}
        & \textcolor{\attcolor}{22.1}  \\
      \cmidrule(lr){1-7}
      0.5
        & 28.4 & 52.9 & 27.0
        & 24.8 & 49.7 & 24.7  \\
      \cmidrule(lr){1-7}
      2.0
        & 29.7 & 54.6 & 28.3
        & 26.9 & 50.5 & 25.3 \\
      \cmidrule(lr){1-7}
      \textcolor{\attcolor}{$\infty$ (only width)}
        & \textcolor{\attcolor}{27.6}
        & \textcolor{\attcolor}{54.6}
        & \textcolor{\attcolor}{27.2}
        & \textcolor{\attcolor}{26.3}
        & \textcolor{\attcolor}{51.1}
        & \textcolor{\attcolor}{25.7}  \\
      \cmidrule(lr){1-7}
      1.0
        & 33.1 & 59.2 & 31.9
        & 31.8 & 56.6 & 30.5 \\
      \bottomrule
 \end{tabular}}
\end{table}

\textbf{Impact of minimum received ratio $\mu$ and waiting interval $T_{clk}$.} 
\textcolor{\attcolor}{We study the impact of $\mu$ and $T_{clk}$ on model performance and the resource utilization of the system, as shown in Fig.\ref{semi-async_exp}.
When $\mu=0.125$ and $T_{clk}=0$, the server updates the full model using fully asynchronous aggregation.
And when $\mu=1$ and $T_{clk}=0$, the server updates the full model synchronously.
When increasing $T_{clk}$ or increasing $\mu$, both the convergence performances of the full model and submodels consistently improve, but the idle computation interval increases, leading to under-utilization of resources.
However, using different $\mu$ and $T_{clk}$ combination can still ensure that $Co\text{-}S^2P$ achieve the stable and better convergence performance and resource utilization compared to the SOTAs.}

\textbf{Impact of weight term $\lambda_1$ to constrain submodel capacity.} 
\textcolor{\attcolor}{We further investigate how the pruning regularization weight $\lambda_1$ affects both global‐model convergence and client‐side accuracy, as shown in Fig.\ref{fig:lambda1}. 
As $\lambda_1$ increases, it enforces a stronger sparsity on the structured-pruned submodel, accelerating the convergence of the submodel. 
Furthermore, it is noted that the smaller $\lambda_1$ does not necessarily lead to better performance. This is because the smaller $\lambda_1$ prevents the client from obtaining a personalized mask adapted to the local data distribution. Moreover, the pruning weight term is stable across a practical range and $\lambda_1=1.0$ provides the best balance between enforcing sparsity and maintaining classification performance.}

\textbf{Impact of balance term $\lambda_2$.} 
\textcolor{\attcolor}{We investigate the impact of $\lambda_2$ on the performances of both the global model and submodels. 
As shown in Fig.\ref{fig:lambda2}, 
When $\lambda_2=0.1, 0.15$, the distillation term is under‑weighted, leading to slower convergence and a lower peak server accuracy. 
At the other extreme ($\lambda_2=0.25,0.3$), over‑emphasis on deep‑block knowledge transfer perturbs the primary classification loss and degrades global performance. 
The intermediate setting $\lambda_2=0.2$ achieves the best trade‑off, yielding the fastest convergence and highest steady‑state accuracy.
Moreover, larger $\lambda_2$ amplifies self‑distillation during submodel training, which accelerates the global model and submodels training. 
These findings confirm that the balance term is robust over a practical range and that $\lambda_2=0.2$ provides an optimal balance between deep‑block knowledge transfer and the classification loss.}

\textbf{Impact of the ratio of $R_n^{width}$ and $R_n^{depth}$.} 
\textcolor{\attcolor}{
We study the impact of different ratio of $R_n^{width}$ and $R_n^{depth}$ on the performance of global model and submodels. 
The ratio indicates the proportion of non-pruned parameters of the submodels along the width and depth dimensions. 
Pure depth pruning (${R_n^{width}}/{R_n^{depth}} = 0$) yields only 23.8\%/22.7\% server/client \textit{Top1} accuracy, while pure width pruning (${R_n^{width}}/{R_n^{depth}} = \infty$) recovers to 27.6\%/26.3\%. These demonstrate that the width structured pruning is more effective in obtaining personalized submodels and final convergence performance.
Furthermore, as shown in Tab.\ref{table:pruneratio}, ${R_n^{width}}/{R_n^{depth}} = 1$ outperforms other ratios, demonstrating that the balanced prune rate enable submodels to have superior learning capacity.}

\begin{table}[t]
  \caption{\textcolor{\attcolor}{Performance comparison on (Bert-Small)-Agnews. Bold is the optimal result except for FedAvg and FedAsync with full model training.}}
  \label{exp:bert-sst2}
  \centering
  \renewcommand{\arraystretch}{0.6}
  \arrayrulecolor{\attcolor}
  \scalebox{0.85}{%
    \begin{tabular}{cccccc}
      \toprule
      \multirow{2}{*}{\textcolor{\attcolor}{Methods}}
        & \multicolumn{2}{c}{\textcolor{\attcolor}{Server}}
        & \multicolumn{2}{c}{\textcolor{\attcolor}{Client(Avg.)}}
        & \multirow{2}{*}{\textcolor{\attcolor}{$RU(\%)$}} \\
      \cmidrule(lr){2-3}\cmidrule(lr){4-5}
      & \textcolor{\attcolor}{$Top1(\%)$}
      & \textcolor{\attcolor}{$F1(\%)$}
      & \textcolor{\attcolor}{$Top1(\%)$}
      & \textcolor{\attcolor}{$F1(\%)$}
      & \\
      \midrule
      \textcolor{\attcolor}{FedAvg}
        & \textcolor{\attcolor}{83.1}
        & \textcolor{\attcolor}{82.7}
        & \textcolor{\attcolor}{82.1}
        & \textcolor{\attcolor}{81.7}
        & \textcolor{\attcolor}{68.2} \\
      \cmidrule(lr){1-6}
      \textcolor{\attcolor}{FedAvg+PLATON}
        & \textcolor{\attcolor}{61.4}
        & \textcolor{\attcolor}{60.1}
        & \textcolor{\attcolor}{58.6}
        & \textcolor{\attcolor}{58.2}
        & \textcolor{\attcolor}{72.9} \\
      \cmidrule(lr){2-6}
      \textcolor{\attcolor}{FedAsync+PLATON}
        & \textcolor{\attcolor}{54.6}
        & \textcolor{\attcolor}{53.9}
        & \textcolor{\attcolor}{53.2}
        & \textcolor{\attcolor}{52.8}
        & \textbf{\textcolor{\attcolor}{100.0}} \\
      \cmidrule(lr){2-6}
      \textcolor{\attcolor}{FedMeZO}
        & \textcolor{\attcolor}{60.1}
        & \textcolor{\attcolor}{59.0}
        & \textcolor{\attcolor}{58.8}
        & \textcolor{\attcolor}{58.4}
        & \textcolor{\attcolor}{70.3} \\
      \cmidrule(lr){2-6}
      \textcolor{\attcolor}{FedRolex}
        & \textcolor{\attcolor}{67.1}
        & \textcolor{\attcolor}{66.8}
        & \textcolor{\attcolor}{66.1}
        & \textcolor{\attcolor}{65.4}
        & \textcolor{\attcolor}{82.4} \\
      \cmidrule(lr){1-6}
      \textcolor{\attcolor}{$\text{Co-S}^2\text{P}$}
        & \textbf{\textcolor{\attcolor}{73.4}}
        & \textbf{\textcolor{\attcolor}{72.9}}
        & \textbf{\textcolor{\attcolor}{72.9}}
        & \textbf{\textcolor{\attcolor}{72.3}}
        & \textcolor{\attcolor}{93.4} \\
      \cmidrule(lr){2-6}
      \textcolor{\attcolor}{$\text{Co-S}^2\text{P}$(Async.)}
        & \textcolor{\attcolor}{70.3}
        & \textcolor{\attcolor}{69.5}
        & \textcolor{\attcolor}{69.6}
        & \textcolor{\attcolor}{68.8}
        & \textbf{\textcolor{\attcolor}{100.0}} \\
      \bottomrule
    \end{tabular}%
  }
\end{table}

\begin{table}[t]
  \caption{\textcolor{\attcolor}{Performance comparison on (ViT-Tiny/16)-ImageNet200. Bold is the optimal result except for FedAvg and FedAsync with full model training.}}
  \label{tab:iotdevices}
  \centering
  \renewcommand{\arraystretch}{0.6}
  \arrayrulecolor{\attcolor}
  \scalebox{0.85}{%
    \begin{tabular}{cccccc}
      \toprule
      \multirow{2}{*}{\textcolor{\attcolor}{Methods}}
        & \multicolumn{2}{c}{\textcolor{\attcolor}{Server}}
        & \multicolumn{2}{c}{\textcolor{\attcolor}{Client(Avg.)}}
        & \multirow{2}{*}{\textcolor{\attcolor}{$RU(\%)$}} \\
      \cmidrule(lr){2-3}\cmidrule(lr){4-5}
      & \textcolor{\attcolor}{$Top1(\%)$}
      & \textcolor{\attcolor}{$F1(\%)$}
      & \textcolor{\attcolor}{$Top1(\%)$}
      & \textcolor{\attcolor}{$F1(\%)$}
      & \\
      \midrule
      \textcolor{\attcolor}{FedAvg}
        & \textcolor{\attcolor}{36.9}
        & \textcolor{\attcolor}{36.4}
        & \textcolor{\attcolor}{35.3}
        & \textcolor{\attcolor}{34.9}
        & \textcolor{\attcolor}{57.2} \\
      \cmidrule(lr){1-6}
      \textcolor{\attcolor}{FedAvg+PLATON}
        & \textcolor{\attcolor}{18.7}
        & \textcolor{\attcolor}{18.1}
        & \textcolor{\attcolor}{17.2}
        & \textcolor{\attcolor}{17.0}
        & \textcolor{\attcolor}{64.1} \\ 
      \cmidrule(lr){2-6}
      \textcolor{\attcolor}{FedAsync+PLATON}
        & \textcolor{\attcolor}{14.2}
        & \textcolor{\attcolor}{14.0}
        & \textcolor{\attcolor}{13.3}
        & \textcolor{\attcolor}{13.0}
        & \textbf{\textcolor{\attcolor}{100.0}} \\ 
      \cmidrule(lr){2-6}
      \textcolor{\attcolor}{FedMeZO}
        & \textcolor{\attcolor}{14.6}
        & \textcolor{\attcolor}{14.4}
        & \textcolor{\attcolor}{13.1}
        & \textcolor{\attcolor}{13.0}
        & \textcolor{\attcolor}{62.4} \\ 
      \cmidrule(lr){2-6}
      \textcolor{\attcolor}{FedRolex}
        & \textcolor{\attcolor}{22.6}
        & \textcolor{\attcolor}{22.3}
        & \textcolor{\attcolor}{19.9}
        & \textcolor{\attcolor}{19.3}
        & \textcolor{\attcolor}{68.4} \\ 
      \cmidrule(lr){1-6}
      \textcolor{\attcolor}{$\text{Co-S}^2\text{P}$}
        & \textbf{\textcolor{\attcolor}{32.5}}
        & \textbf{\textcolor{\attcolor}{32.0}}
        & \textbf{\textcolor{\attcolor}{31.6}}
        & \textbf{\textcolor{\attcolor}{31.2}}
        & \textcolor{\attcolor}{84.5} \\
      \cmidrule(lr){2-6}
      \textcolor{\attcolor}{$\text{Co-S}^2\text{P}$(Async.)}
        & \textcolor{\attcolor}{26.9}
        & \textcolor{\attcolor}{26.4}
        & \textcolor{\attcolor}{26.1}
        & \textcolor{\attcolor}{25.8}
        & \textbf{\textcolor{\attcolor}{100.0}} \\
      \bottomrule
    \end{tabular}%
  }
\end{table}

\begin{table}[t]
  \caption{\textcolor{\attcolor}{Performance comparison on (LLaMA-Tiny)-Fingpt.}}
  \label{tab:scalemodel}
  \centering
  \renewcommand{\arraystretch}{0.58}
  \arrayrulecolor{\attcolor}
  \scalebox{0.7}{%
    \begin{tabular}{ccccccc}
      \toprule
      \multirow{2}{*}{\textcolor{\attcolor}{Methods}}
        & \multicolumn{2}{c}{\textcolor{\attcolor}{Server}}
        & \multicolumn{2}{c}{\textcolor{\attcolor}{Client(Avg.)}}
        & \multirow{2}{*}{\textcolor{\attcolor}{$RU(\%)$}}& \multirow{2}{*}{\textcolor{\attcolor}{$Comm.(GB)$}} \\
      \cmidrule(lr){2-3}\cmidrule(lr){4-5}
      & \textcolor{\attcolor}{$Top1(\%)$}
      & \textcolor{\attcolor}{$F1(\%)$}
      & \textcolor{\attcolor}{$Top1(\%)$}
      & \textcolor{\attcolor}{$F1(\%)$}
      & &\\
      \midrule
      \textcolor{\attcolor}{FedAvg}
        & \textcolor{\attcolor}{71.7}
        & \textcolor{\attcolor}{71.4}
        & \textcolor{\attcolor}{69.5}
        & \textcolor{\attcolor}{69.0}
        & \textcolor{\attcolor}{62.7}
        & \textcolor{\attcolor}{30.7}\\
      \cmidrule(lr){1-7}
      \textcolor{\attcolor}{FedAvg+PLATON}
        & \textcolor{\attcolor}{49.8}
        & \textcolor{\attcolor}{49.3}
        & \textcolor{\attcolor}{48.7}
        & \textcolor{\attcolor}{48.2}
        & \textcolor{\attcolor}{73.6} 
        & \textcolor{\attcolor}{20.1}\\ 
      \cmidrule(lr){2-7}
      \textcolor{\attcolor}{FedAsync+PLATON}
        & \textcolor{\attcolor}{42.1}
        & \textcolor{\attcolor}{41.7}
        & \textcolor{\attcolor}{41.0}
        & \textcolor{\attcolor}{40.7}
        & \textbf{\textcolor{\attcolor}{100.0}} 
        & \textcolor{\attcolor}{1.4($\pm0.74$)}\\ 
      \cmidrule(lr){2-7}
      \textcolor{\attcolor}{FedRolex}
        & \textcolor{\attcolor}{53.7}
        & \textcolor{\attcolor}{53.4}
        & \textcolor{\attcolor}{51.8}
        & \textcolor{\attcolor}{51.3}
        & \textcolor{\attcolor}{80.2} 
        & \textcolor{\attcolor}{19.6}\\
      \cmidrule(lr){1-7}
      \textcolor{\attcolor}{$\text{Co-S}^2\text{P}$}
        & \textbf{\textcolor{\attcolor}{60.2}}
        & \textbf{\textcolor{\attcolor}{59.8}}
        & \textbf{\textcolor{\attcolor}{59.4}}
        & \textbf{\textcolor{\attcolor}{59.0}}
        & \textbf{\textcolor{\attcolor}{91.5}} 
        & \textcolor{\attcolor}{11.3($\pm3.7$)}\\
      \cmidrule(lr){2-7}
      \textcolor{\attcolor}{$\text{Co-S}^2\text{P}$(Async.)}
        & \textcolor{\attcolor}{56.7}
        & \textcolor{\attcolor}{56.2}
        & \textcolor{\attcolor}{55.4}
        & \textcolor{\attcolor}{55.2}
        & \textbf{\textcolor{\attcolor}{100.0}} 
        & \textcolor{\attcolor}{0.82($\pm0.61$)}\\
      \bottomrule
    \end{tabular}%
  }
\end{table}


\subsection{Scalability Analysis}

\textcolor{\attcolor}{\textbf{Scaling to NLP task.} 
We extend the proposed framework $Co\text{-}S^2P$ to NLP task, training the BERT-Small~\cite{devlin2019bert} on the SST-2 sentiment classification dataset. 
The real-world testbed setting follows the experiment setting of E1, as shown in Tab.\ref{exp:bert-sst2}.
Compared to the SOTA FedRolex, $Co\text{-}S^2P$ improves server-side \textit{Top1} accuracy by 6.3\% and client-side average \textit{Top1} accuracy by 6.8\%, while improving overall RU by 11.0\%. 
Additionally, $Co\text{-}S^2P$(Async.) maintains higher convergence performance and the best resource utilization compared to the baselines.
These results confirm that $Co\text{-}S^2P$ can efficiently scale to different training tasks and achieve better convergence performance and resource utilization.}

\textcolor{\attcolor}{\textbf{Scaling various IoT devices.}
In terms of hardware diversity, we expand our real-world testbed to include not only various Jetson devices (AGX Orin and Xavier NX) but also Raspberry Pi and laptop devices. 
The detailed training setup follows the experiment setting of E1.
As shown in Tab.\ref{tab:iotdevices}, $Co\text{-}S^2P$ improves server-side \textit{Top1} accuracy by 6.3\% and client-side average \textit{Top1} accuracy by 6.8\%.
Additionally, $Co\text{-}S^2P$(Async.) achieves 25.6\% \textit{Top1} accuracy and substantially outperforms the baselines. 
These results demonstrate that $Co\text{-}S^2P$ effectively leverages heterogeneous computing and memory resources and achieve robust convergence performance across various heterogeneous IoT devices.}

\begin{figure}[t]
  \centering
  \includegraphics[width=0.48\textwidth]{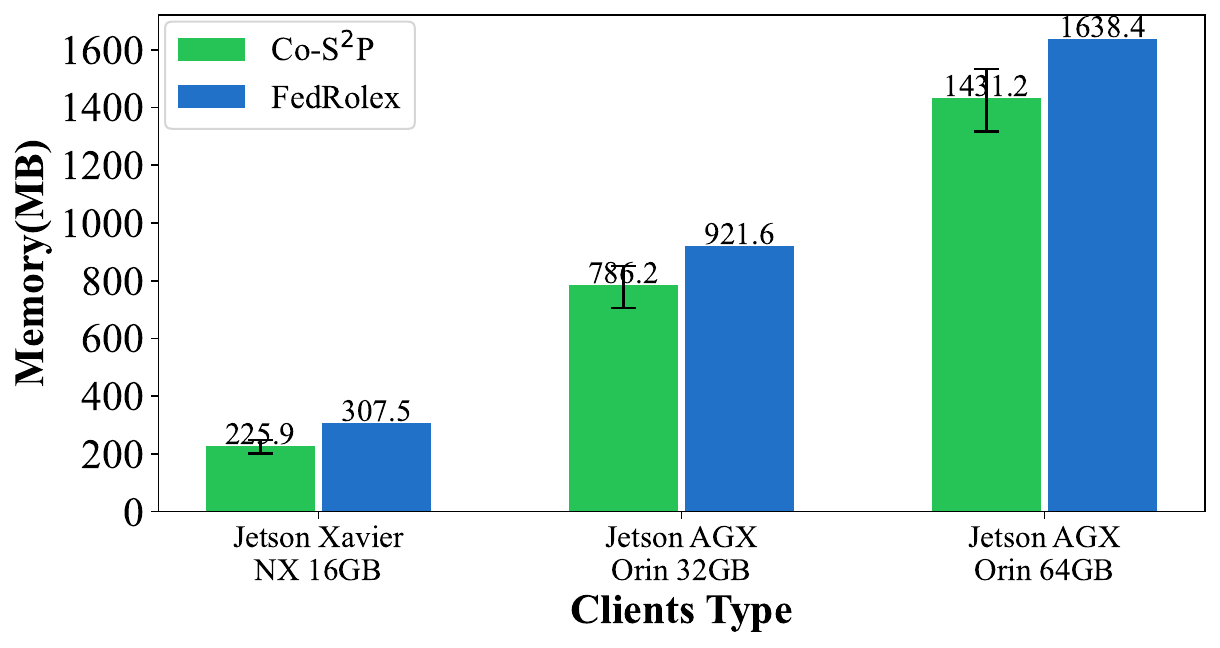}
  \caption{\textcolor{\attcolor}{Comparison of model memory usage during full parameter fine-tuning of the llama-tiny model (1.1B).}}
  \label{exp:llamamem}
\end{figure}

\textcolor{\attcolor}{\textbf{Scaling to larger scale model.} 
In the experiment, we full parameter fine-tune the LLaMA-tiny model~\cite{touvron2023llama} with parameters up to 1.1B using the Fingpt~\cite{yang2023fingpt} and evaluate it using the FPB benchmark~\cite{malo2014good}, as shown in Tab.\ref{tab:scalemodel} and Fig.\ref{exp:llamamem}.
Compared to FedRolex, $Co\text{-}S^2P$ improves server-side \textit{Top1} accuracy by 6.5\% and client-side average \textit{Top1} accuracy by 7.6\%, while achieving a significantly higher resource utilization rate of 91.5\%. 
Additionally, $Co\text{-}S^2P$ reduces communication overhead per round by about 42.3\% and memory overhead by about 18.5\%.
The experimental results demonstrate that $Co\text{-}S^2P$ effectively scales to large-scale models training in heterogeneous and resource-constrained scenarios.}

\section{Conclusion}
To release the potential of massive resource-limited nodes, we proposed a novel semi-asynchronous collaborative training framework ${Co\text{-}S}^2{P}$. 
We designed a data distribution-aware structured pruning to ensure balanced learning capability and accelerate training.
By self-distillation, ${Co\text{-}S}^2{P}$ facilitated shallow blocks obtaining the high-level knowledge from the deep blocks.
Furthermore, we introduced a semi-asynchronous aggregation strategy to mitigate the straggler problem, and theoretically proved the strategy converges with $O(1/\sqrt{N^*EQ})$.
We conducted the real-world experiments training model with parameters up to 0.11B. 
${Co\text{-}S}^2{P}$ outperforms all the baselines, while reducing memory consumption and training time of all clients and improving resource utilization of the system across different types of tasks.






\bibliographystyle{IEEEtran}
\bibliography{example_paper}

\vspace{11pt}


\begin{IEEEbiography}[{\includegraphics[width=1in,height=1.25in,clip,keepaspectratio]{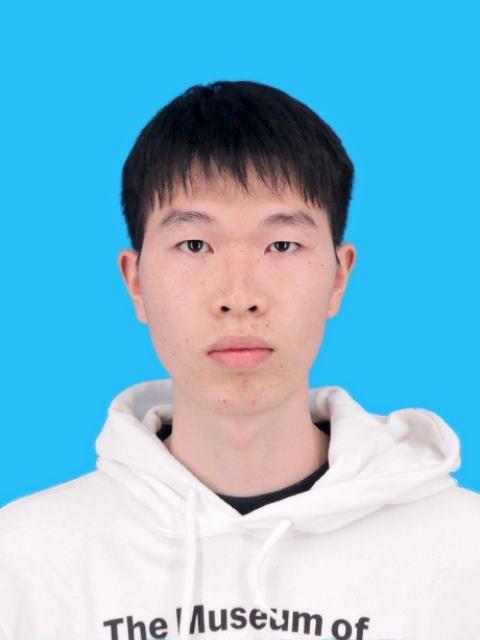}}]{Yan Li} is currently pursuing the M.S. degree with the School of Computer Science and Technology, Shandong University. He received his B.S. degree in Shandong University. His research interests include collaborative learning and distributed systems.
\end{IEEEbiography}
\vspace{-30pt}
\begin{IEEEbiography}[{\includegraphics[width=1in,height=1.25in,clip,keepaspectratio]{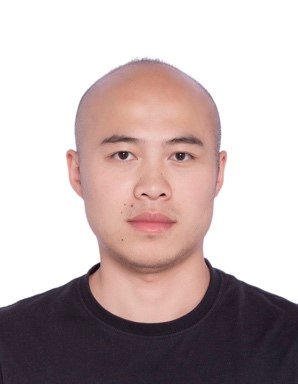}}]{Xiao Zhang} is now an associate professor in the School of Computer Science and Technology, Shandong University. His research interests include data mining, multi-task learning and federated learning. 
He has published more than 20 papers in the prestigious refereed journals and conference proceedings, such as IEEE Transactions on Mobile Computing, UBICOMP, ACM MULTIMEDIA,  IJCAI, AAAI, ACM CIKM, and IEEE ICDM. 
\end{IEEEbiography}
\vspace{-30pt}
\begin{IEEEbiography}[{\includegraphics[width=1in,height=1.25in,clip,keepaspectratio]{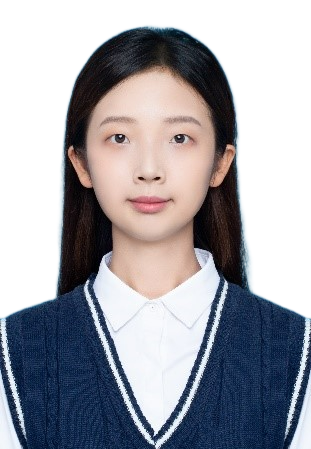}}]{Mingyi Li} is currently a PhD student in the School of Computer Science and Technology, Shandong University. She received her B.S. degree in Shandong University. Her research interests include distributed collaborative Learning and the theoretical optimization of distributed algorithms.
\end{IEEEbiography}
\vspace{-30pt}
\begin{IEEEbiography}[{\includegraphics[width=1in,height=1.25in,clip,keepaspectratio]{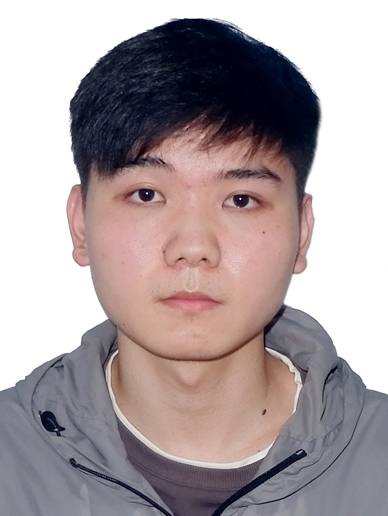}}]{Guangwei Xu} is currently pursuing the M.S. degree with the School of Computer Science and Technology, Shandong University. He received his B.S. degree in Shandong University. His research interests include federated learning and machine learning.
\end{IEEEbiography}
\vspace{-30pt}
\begin{IEEEbiography}[{\includegraphics[width=1in,height=1.25in,clip,keepaspectratio]{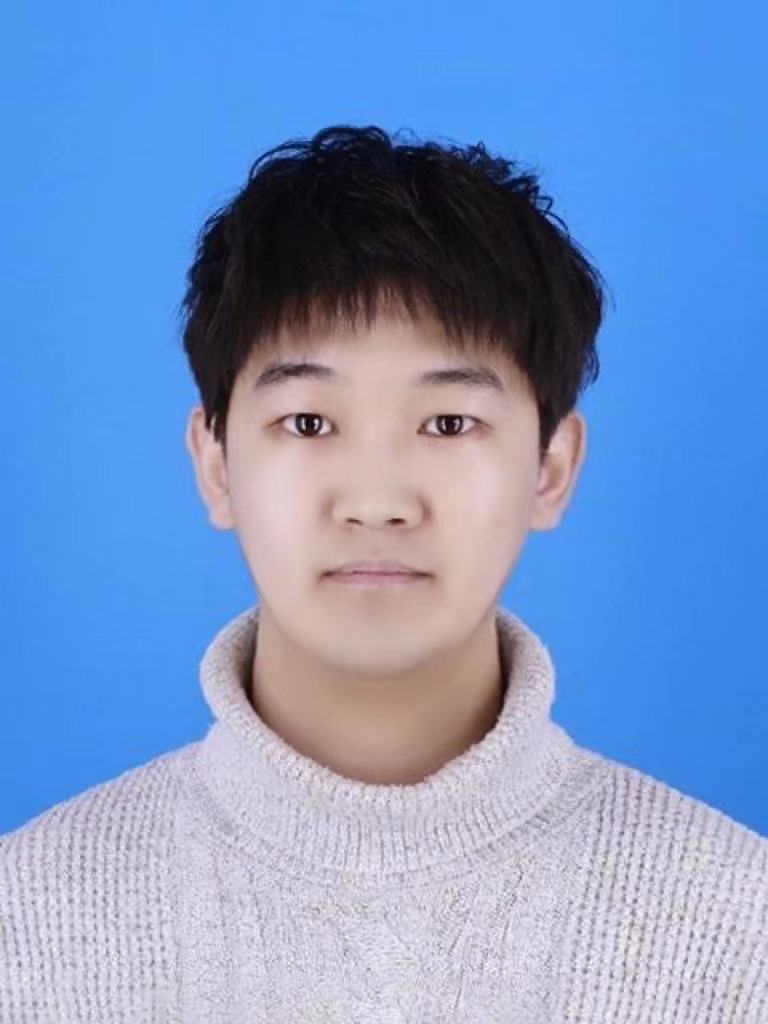}}]{Feng Chen} is currently pursuing his Ph.D. at the School of Computer Science and Technology, Shandong University, where he also completed his M.S. degree. His research focuses on deep neural networks and GPU acceleration algorithms.
\end{IEEEbiography}
\vspace{-30pt}
\begin{IEEEbiography}[{\includegraphics[width=1in,height=1.25in,clip,keepaspectratio]{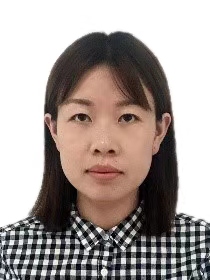}}]{Yuan Yuan} received the BSc degrees from the School of Mathematical Sciences, Shanxi University in 2016, and the Ph.D. degree from the School of  Computer Science and Technology, Shandong University, Qingdao, China, in 2021. She is currently a postdoctoral fellow at the Shandong University-Nanyang Technological University International Joint Research Institute on Artificial Intelligence, Shandong University. Her research interests include distributed computing and distributed machine learning.
\end{IEEEbiography}
\vspace{-30pt}
\begin{IEEEbiography}[{\includegraphics[width=1in,height=1.25in,clip,keepaspectratio]{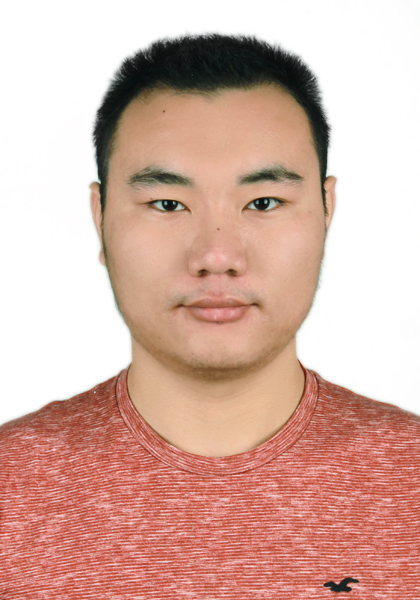}}]{Yifei Zou} received the B.E. degree in 2016 from Computer School, Wuhan University, and the PhD degree in 2020 from the Department of Computer Science, The University of Hong Kong. He is currently an Assistant Professor with the school  of computer science and technology, Shandong University. His research interests include wireless networks, ad hoc networks and distributed computing.
\end{IEEEbiography}
\vspace{-30pt}
\begin{IEEEbiography}[{\includegraphics[width=1in,height=1.25in,clip,keepaspectratio]{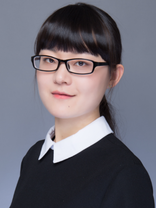}}]{Mengying Zhao} received B.E. degree in School of
Computer Science and Technology from Shandong
University, China in July 2011, and Ph.D. degree in
Department of Computer Science, City University
of Hong Kong in July 2015. She is now a professor
in School of Computer Science and Technology at
Shandong University. Her research interests include
computer architecture, embedded system, and nonvolatile memory.
\end{IEEEbiography}
\vspace{-30pt}
\begin{IEEEbiography}[{\includegraphics[width=1in,height=1.25in,clip,keepaspectratio]{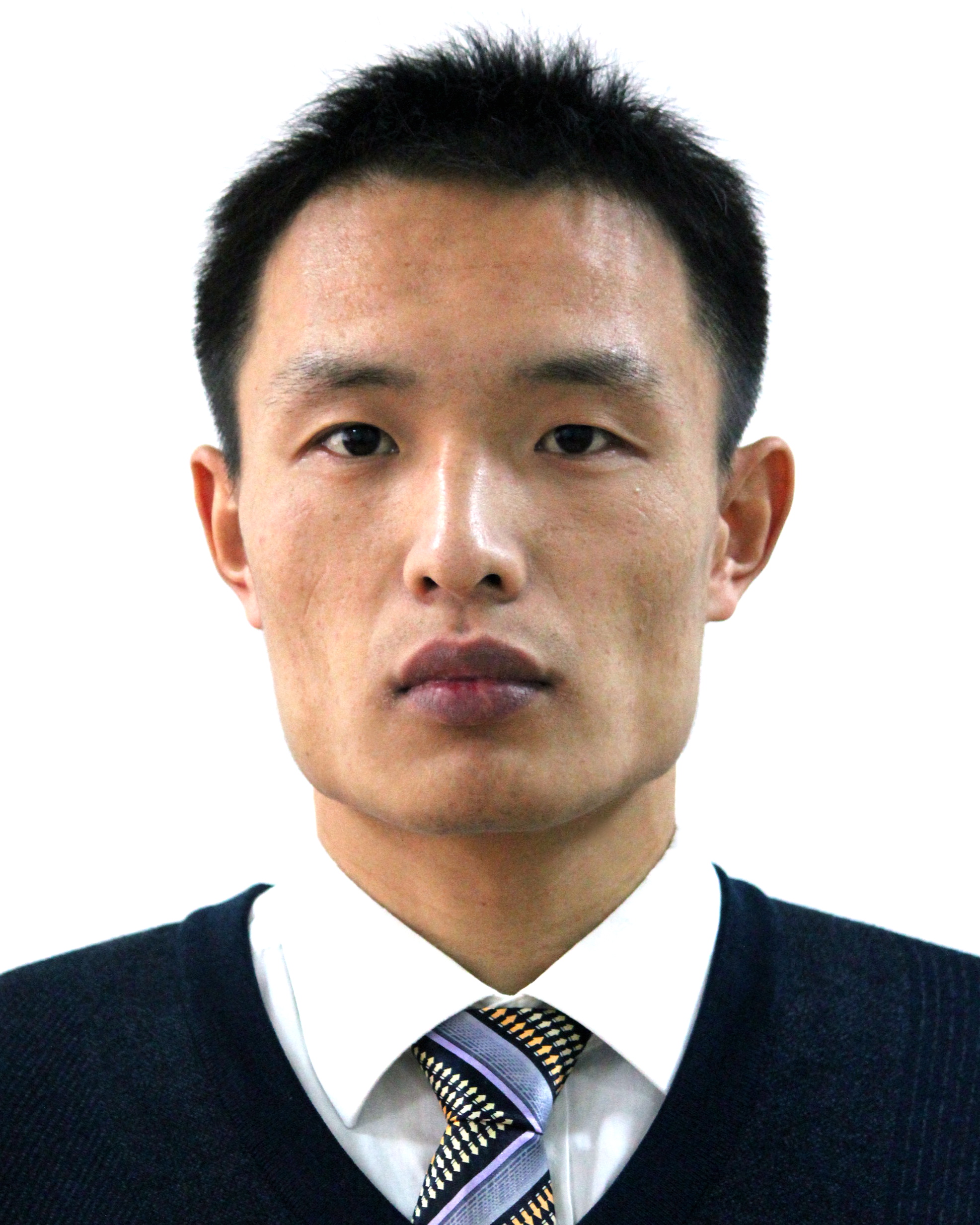}}]{Jianbo Lu} received the Ph.D. degree in Department of Automation, Shanghai Jiao Tong University in 2015. He is now an associate professor with School of Computer Science and Technology at Shandong University. His research interests include federated learning, control and optimization, and model predictive control.
\end{IEEEbiography}
\vspace{-30pt}
\begin{IEEEbiography}[{\includegraphics[width=1in,height=1.25in,clip,keepaspectratio]{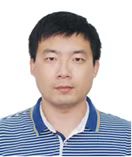}}]{Dongxiao Yu} received the B.S. degree in 2006 from the School of Mathematics, Shandong University and the Ph.D degree in 2014 from the Department of Computer Science, The University of Hong Kong. He became an associate professor in the School of Computer Science and Technology, Huazhong University of Science and Technology, in 2016. He is currently a professor in the School of Computer Science and Technology, Shandong University. His research interests include edge intelligence, distributed computing and data mining.
\end{IEEEbiography}



\vfill

\end{document}